\documentclass[paper]{agujournal2019}
\usepackage{url}
\usepackage{lineno}
\usepackage{color}
\usepackage{amssymb,amsmath}
\usepackage{setspace}
\draftfalse

\newcommand{\ssr}{   {Space Sci. Rev. }}

\newcommand{\jgr}{   {J. Geophys. Res.}}
\newcommand{\grl}{   {Geophys. Res. Lett.}}

\newcommand{\apjl}{   {Astrophys. J. Lett.}}

\newcommand{\nat}{   {Nature}}
\newcommand{\aap}{   {Astronomy \& Astrophysics}}

\journalname{JGR: Space Physics}

\begin{document}

%% ---------------------------------------------------------------%%

\title{Equatorial source of oblique electromagnetic ion cyclotron waves: peculiarities in the ion distribution function}

\authors{David S. Tonoian\affil{1}, Xiao-Jia Zhang \affil{1,2}, Anton Artemyev\affil{2}, Xin An\affil{2}} %Qianli? maybe useful to him if he has his EMIC project
\affiliation{1}{Department of Physics, University of Texas at Dallas, Richardson, TX, USA}
\affiliation{2}{Department of Earth, Planetary, and Space Sciences, University of California, Los Angeles, USA}

\correspondingauthor{David S. Tonoian}{david.tonoian@utdallas.edu}

\begin{keypoints}
\item We report observations of very oblique EMIC waves around their equatorial source region
\item Oblique EMIC wave generation is associated with field-aligned thermal ion streams and hot, transversely anisotropic ions
\item The presence of field-aligned thermal ion streams attributes oblique EMIC wave generation to its possible ionospheric source
\end{keypoints}

\begin{abstract}
Electromagnetic ion cyclotron (EMIC) waves are important for Earth’s inner magnetosphere as they can effectively drive relativistic electron losses to the atmosphere and energetic (ring current) ion scattering and isotropization. EMIC waves are generated by transversely anisotropic ion populations around the equatorial source region, and for typical magnetospheric conditions this almost always produces field-aligned waves. For many specific occasions, however, oblique EMIC waves are observed, and such obliquity has been commonly attributed to the wave off-equatorial propagation in curved dipole magnetic fields. In this study, we report that very oblique EMIC waves can be directly generated at the equatorial source region. Using THEMIS spacecraft observations at the dawn flank, we show that such oblique wave generation is possible in the presence of a field-aligned thermal ion population, likely of ionospheric origin, which can reduce Landau damping of oblique EMIC waves and cyclotron generation of field-aligned waves. This generation mechanism underlines the importance of magnetosphere-ionosphere coupling processes in controlling wave characteristics in the inner magnetosphere.
\end{abstract}

\section{Introduction}
Electromagnetic ion cyclotron (EMIC) waves is a natural emission generated by transversely anisotropic ion (proton) populations in the Earth’s inner magnetosphere \cite{Cornwall70,Cornwall&Schulz71,Min15:emic,Yue19:emic}. These waves are responsible for resonant scattering of relativistic electrons and energetic (ring current) ions \cite<see review>[and references therein]{Usanova&Mann16}, which makes this wave mode principally important for the inner magnetosphere dynamics. 

Transversely anisotropic ion populations, either injected from the plasma sheet or formed due to the magnetosphere compression by solar wind transients \cite{Jun19:emic,Jun21:emic,Kim21:emic}, usually generate field-aligned EMIC waves \cite{Chen10:emic,Chen11:emic}, whereas generation of oblique waves is often suppressed by wave Landau damping due to suprathermal ions \cite<see discussions in>{SoriaSantacruz13}. Thus, oblique EMIC waves have been mostly detected off the equator \cite{Liu13:emic}, because wave propagation in highly inhomogeneous magnetic field and plasma will result in wavevector deviation from the field-aligned direction \cite{Rauch&Roux82,Thorne&Horne97,Fraser&Nguyen01}. It has been shown that oblique EMIC waves can scatter particles via several resonant mechanisms not available to field-aligned waves \cite<see review by>{Usanova21}. First, wave obliquity modifies the efficiency of relativistic electron scattering \cite<see>{Khazanov&Gamayunov07,Lee18:emic,Hanzelka23}. Second, the Landau resonance between oblique EMIC waves and cold ions allows EMIC waves to serve as a transmitter of energy between hot and cold ion populations \cite{Omura85,Kitamura18,Ma19}. Third, oblique EMIC waves can resonate with cold plasmasphere electrons \cite<in linear and nonlinear regimes, see>{Wang19:emic} and can accelerate and/or precipitate them to form the stable red aurora arcs \cite{Cornwall71:arc,Thorne&Horne92}. Fourth, oblique EMIC waves may scatter energetic ($\sim 100$keV) electrons via the bounce resonance \cite{Wang18:bounce,Blum19}. All these mechanisms imply the importance of understanding possible sources of oblique EMIC waves: can such waves be generated around the equator or only off-equator as the field-aligned waves propagate away from their equatorial source?

The equatorial generation of oblique EMIC waves requires a viable mechanism to suppress the Landau damping, i.e., reduction of the field-aligned gradient in the ion phase space density around the Landau resonant energies (hundreds of eV). For the electron-scale whistler-mode waves, it has been shown that oblique wave generation can be explained by field-aligned electron streams generated by ionosphere outflow \cite<see discussions in>[]{Artemyev&Mourenas20:jgr}, which largely suppress Landau damping \cite<see, e.g.,>{Mourenas15,Li16}. Such ionosphere outflows also include field-aligned ionospheric ion populations, which are often detected in the inner magnetosphere and near-Earth plasma sheet \cite{Yue17:ions,Artemyev18:jgr:RBSP&THEMIS}. Therefore, oblique EMIC wave generation may be explained by a combination of near-equatorial, transversely anisotropic hot ions and field-aligned thermal ion streams. 

In this study, we investigate near-equatorial observations of very oblique EMIC waves by Time History of Events and Macroscale Interactions during Substorms (THEMIS) spacecraft \cite{Angelopoulos08:ssr}. Combining measurements of ion distribution functions \cite{McFadden08:THEMIS} and linear dispersion solver \cite{Astfalk&Jenko17}, we reveal the conditions of such wave generation and demonstrate that very oblique EMIC waves are associated with field-aligned ion streams. The rest of this paper is structured as follows: Section~\ref{sec:data} describes THEMIS instrumentation, plasma and wave measurements, methods of data analysis, and details of the linear dispersion solver. Section~\ref{sec:events} examines the generation of multiple very oblique EMIC wave events by analyzing the observed ion distribution function and comparing this distribution with that during field-aligned EMIC waves. Section~\ref{sec:discussion} summarizes our results and discusses their implication for modelling of particle dynamics in the inner magnetosphere.

\section{Dataset and instruments}\label{sec:data}
We combine near-equatorial THEMIS measurements of EMIC waves \cite<of 16Hz sampling rate from the flux-gate magnetometer, see>{Auster08:THEMIS}, plasma sheet ion distributions \cite<at 3s resolution from the electrostatic analyzer (ESA), covering $<25$keV energy range and full pitch-angle range, see>{McFadden08:THEMIS}, total plasma density inferred from the spacecraft potential \cite<see>{Bonnell08,Nishimura13:density}, and
the numerical dispersion solver for electromagnetic waves -- Linear Electromagnetic Oscillations in Plasmas with Arbitrary Rotationally-symmetric Distributions \cite<LEOPARD, see>{Astfalk&Jenko17}. We use the single-spacecraft maximum variance analysis technique \cite{Means72} to estimate the wave normal angle, which comes with an ambiguity of parallel versus anti-parallel directions: the wave normal angle from this technique ranges from $0^\circ$ to $90^\circ$, rather than $0^\circ$ to $180^\circ$.  

An important advantage of LEOPARD model is it can accommodate arbitrary gyrotropic distribution functions in a uniform grid of velocities parallel and perpendicular to the background magnetic field, $(v_\parallel, v_\perp)$. Compared to dispersion solvers that require fitting the measurements to prescribed particle distributions (such as Maxwellians), LEOPARD significantly reduces the uncertainties due to fitting and hence can give a more reliable dispersion solution. Therefore, we use ESA ion distributions in $23$ logarithmically spaced energy channels between $5$ and $25$ keV and $8$ linearly spaced pitch-angle channels and interpolate them to a denser $(v_\parallel, v_\perp)$ grid, which is then passed to LEOPARD to evaluate the observed EMIC wave dispersion and growth rate. We will combine LEOPARD and THEMIS measurements to reveal properties of specific ion distributions that are responsible for the generation of very oblique EMIC waves, quite atypical type of EMIC emission. Note that although THEMIS ESA provides plasma sheet ($<30$keV) electron distributions, this electron population usually does not resonate with EMIC waves or alter the wave dispersion. Thus, we only treat electrons as the cold background plasma (with electron $\beta=10^{-2}$) and do not examine their contribution to EMIC wave growth.

\section{Oblique EMIC events}\label{sec:events}  
We now analyze in details multiple events with very oblique EMIC waves. These are typical events of this wave population and hence their characteristics will be representative of the entire population. 

\subsection{Detailed analysis of the first event}\label{sec:event}
\begin{figure}
        \centering
        \includegraphics[width = 0.8\textwidth]{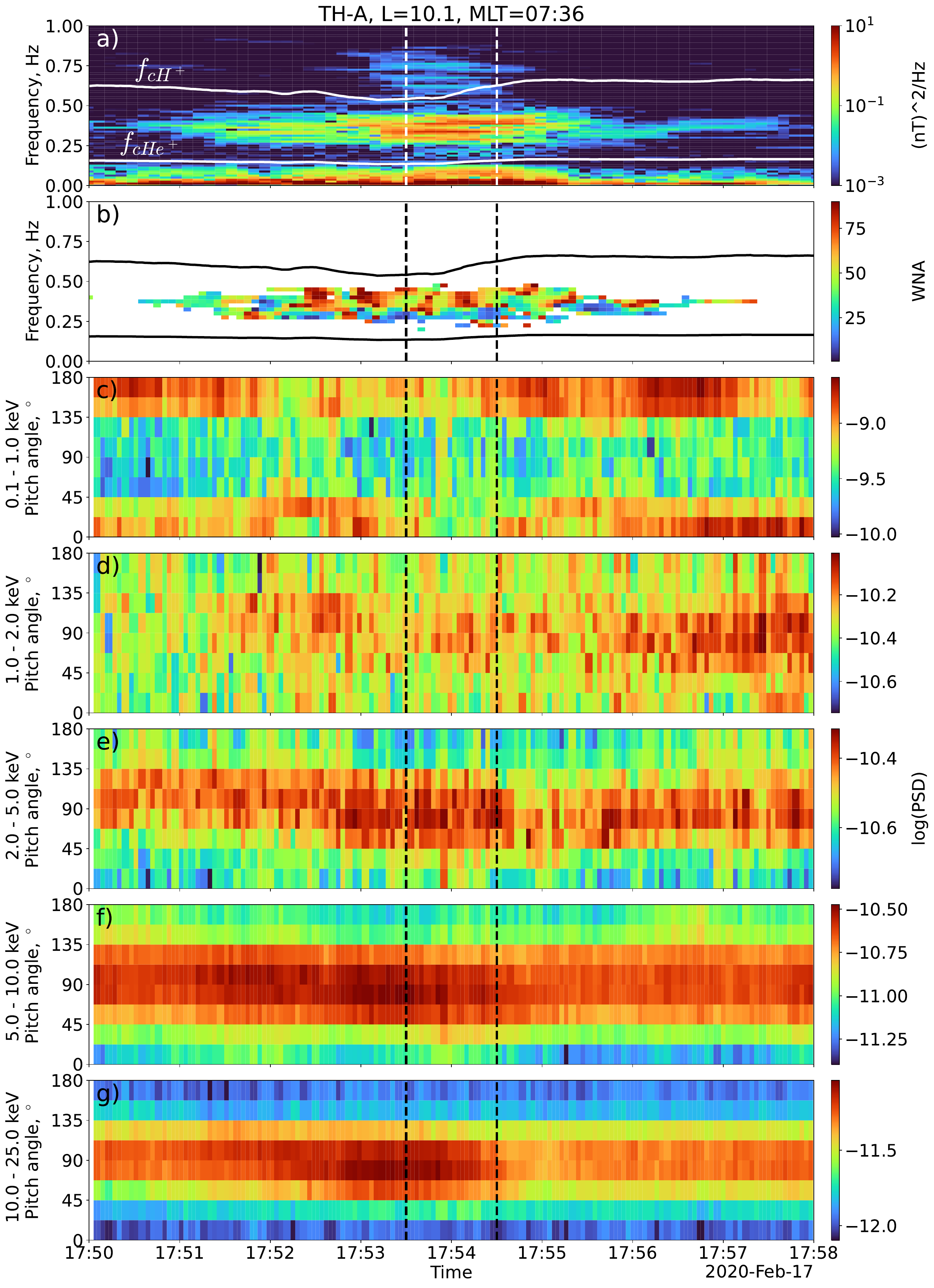}
        \caption{Observation of oblique, hydrogen-band EMIC waves by THEMIS-A spacecraft on February 17, 2020. a) Wave magnetic field power spectrum, b) wave normal angle; top and bottom lines represent hydrogen ion $H^+$ and helium ion $He^+$ cyclotron frequencies, respectively; vertical dashed lines mark the time interval that is used to average the ion phase space density for subsequent investigations of wave dispersion properties. c-g) Ion pitch-angle distributions for different energy ranges as a function of time.}
        \label{fig:2020-02-17spectr}
    \end{figure}

The first event shows very oblique EMIC waves observed by THEMIS-A spacecraft on February 17, 2020, between 17:50-17:58 UT (Figure \ref{fig:2020-02-17spectr}). During this event, hydrogen-band EMIC waves (of frequencies within $0.3-0.9$ of the proton cyclotron frequency $f_{cH+}$) were detected in the dawn flank of the outer edge of the inner magnetosphere (MLT $\sim 07$, $L\sim 10$) with the presence of hot, plasma sheet ions. The EMIC wave normal angle can reach $80^\circ$ during this event. The ion beta is $\beta \sim 1.7$, typical for the inner plasma sheet edge/outer edge of the inner magnetosphere \cite{Yue17:ions,Artemyev18:jgr:RBSP&THEMIS}. 

Figure \ref{fig:2020-02-17spectr} shows the field-aligned population of thermal ($\lesssim 1$ keV) ions and hot ($>2$keV), transversely anisotropic ion population. Both the flux and anisotropy of hot ions are large right around the moment of intense, very oblique EMIC waves. Such transversely anisotropic ions are likely responsible for EMIC wave generation \cite{Chen10:emic,Chen11:emic,Yue19:emic}, but instead of more typical field-aligned waves we observe very oblique waves. Therefore, certain features in the ion distribution significantly alter the generation mechanism. Most likely the thermal field-aligned ion population affects wave generation and moves the positive growth rate to high wave normal angles. To verify this assumption, we will combine the measured ion distribution and linear dispersion solver.

%Most likely the thermal field-aligned ion population reduces the Landau damping of oblique waves and makes it possible for hot transversly anisotropic population to generation such oblique waves. To examine this assumption we will combine measured ion distribution and linear dispersion solver.

\begin{figure*}
        \centering
            \includegraphics[width=.75\textwidth]{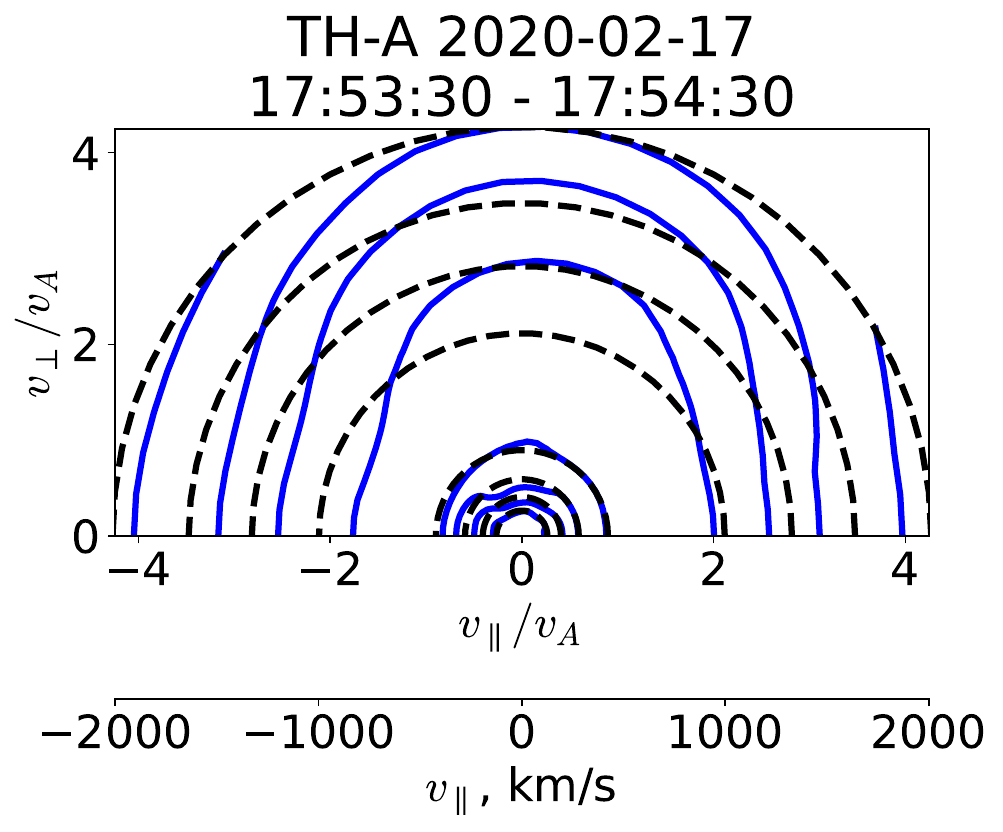}
        \caption{Snapshot of the ion distribution observed by THEMIS-A during the event on February 17, 2020 from Fig. \ref{fig:2020-02-17spectr}; ESA measurements are averaged over 17:53:30-17:54:30 UT, which are used to plot the contours of the ion phase space density (blue). For reference, the black dashed traces show contours of the particle energy.}
        \label{fig:2020-02-17ion_distr}
    \end{figure*}

\begin{figure*}
     \centering
            \includegraphics[width=.75\textwidth]{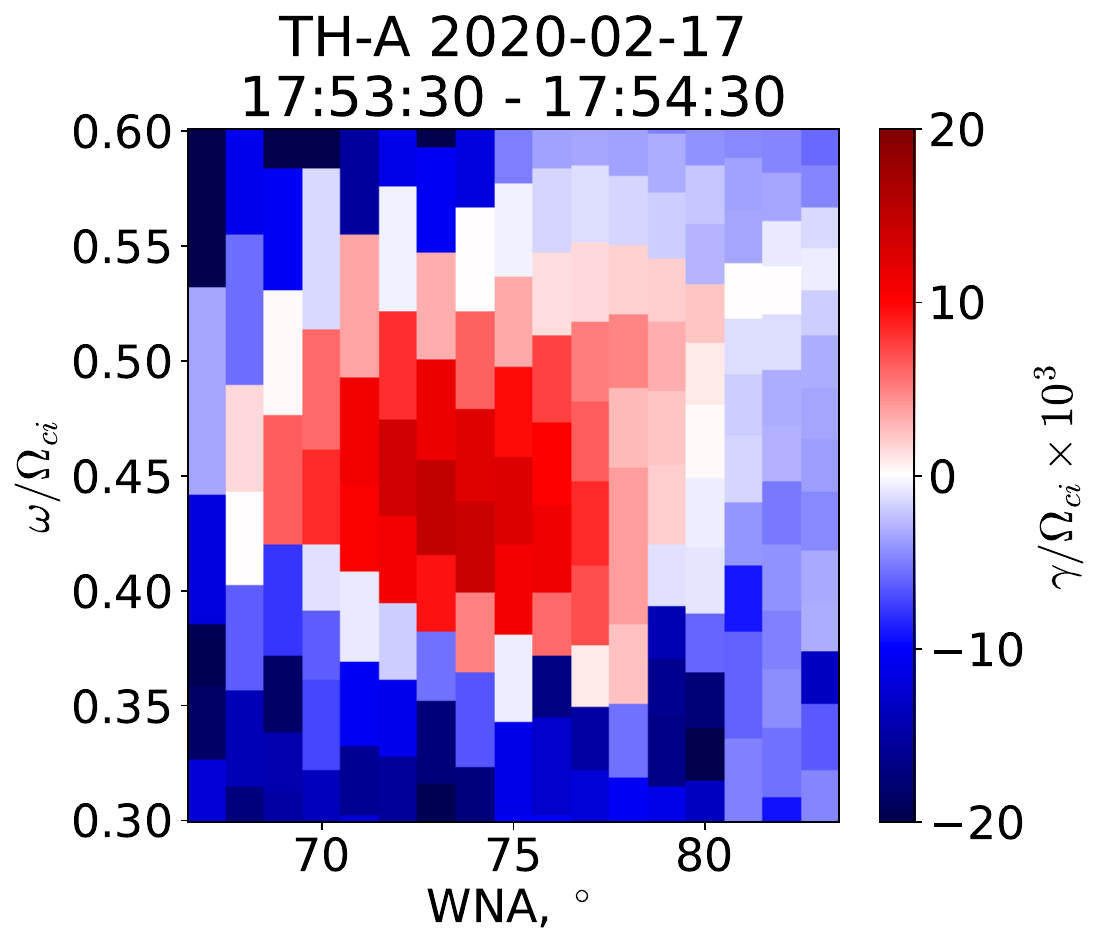}
        \caption{The estimated EMIC wave growth rate in the plane of wave normal angle and wave frequency. The growth rate is evaluated for  hydrogen-electron plasma with the ion distribution from Fig. \ref{fig:2020-02-17ion_distr} .}
        \label{fig:2020-02-17growth_rate}
        
    \end{figure*}

    Both the field-aligned thermal ion population and hot, transversely anisotropic population can be well seen in Figure \ref{fig:2020-02-17ion_distr}, where we plot the 1-min averaged (around the time of the most intense wave spectrum) ion distribution in $(v_\parallel, v_\perp)$ plane. The velocities are normalized to the Alfv\a'en velocity $v_A = B/\sqrt{4\pi n m_p}$, where $n$ is the plasma density calculated from spacecraft potential, $m_p$ is proton mass. Compared with the isotropic distribution (shown in black dashed curves), observations clearly show a strong transverse anisotropy at $v_{\perp}>0.5v_A$ and field-aligned anisotropy at $v_{\perp}<0.25v_A$. This $(v_\parallel, v_\perp)$ distribution is then passed to the LEOPARD solver \cite{Astfalk&Jenko17} to calculate the wave dispersion relation, where we can determine the frequency--wave normal angle region of positive growth rate and compare this with observations of very oblique EMIC waves. Figure \ref{fig:2020-02-17growth_rate} shows results of LEOPARD calculations: the positive growth rate is indeed confined to very oblique wave normal angles, peaking at $74^\circ$, and to wave frequencies around $0.35-0.55 f_{cH+}$ (note throughout the paper we also use $\Omega_{ci}=2\pi f_{cH+}$). These ranges of wave normal angles and frequencies are quite close to those observed by THEMIS (see Fig.\ref{fig:2020-02-17spectr}), which confirms that the ion distribution from Fig. \ref{fig:2020-02-17ion_distr} can reliably reproduce the main wave properties. Note that there is no positive growth rate for field-aligned (or small wave normal angle) waves (not shown).
    
    %Presented wave normal angle range could be extended, to the left but it does not include any more positive growth rate regions. Furthermore, for small propagation angles it becomes hard to produce reliable dispersion solutions, possible explanation for which will be described later in this Section.
    

\begin{figure}
    \centering
    \includegraphics[width=\textwidth]{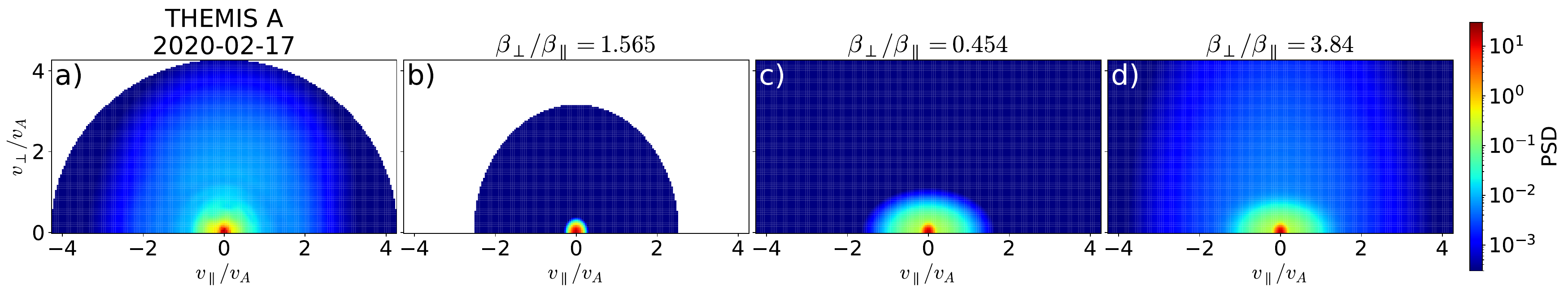}
    \caption{Observed ion distribution during the event on February 17, 2020 (a) and its fitting to a sum of three bi-maxwellian distributions (b-d): panel (b) shows a single cold ion population with $\beta<10^{-2}$, panel (c) shows a summation of cold and thermal ion populations, panel (d) shows a summation of cold, thermal, and hot ion populations.}
    \label{fig:2020-02-17bi-max_fit}
\end{figure}

\begin{figure}
    \centering
    \includegraphics[width=\textwidth]{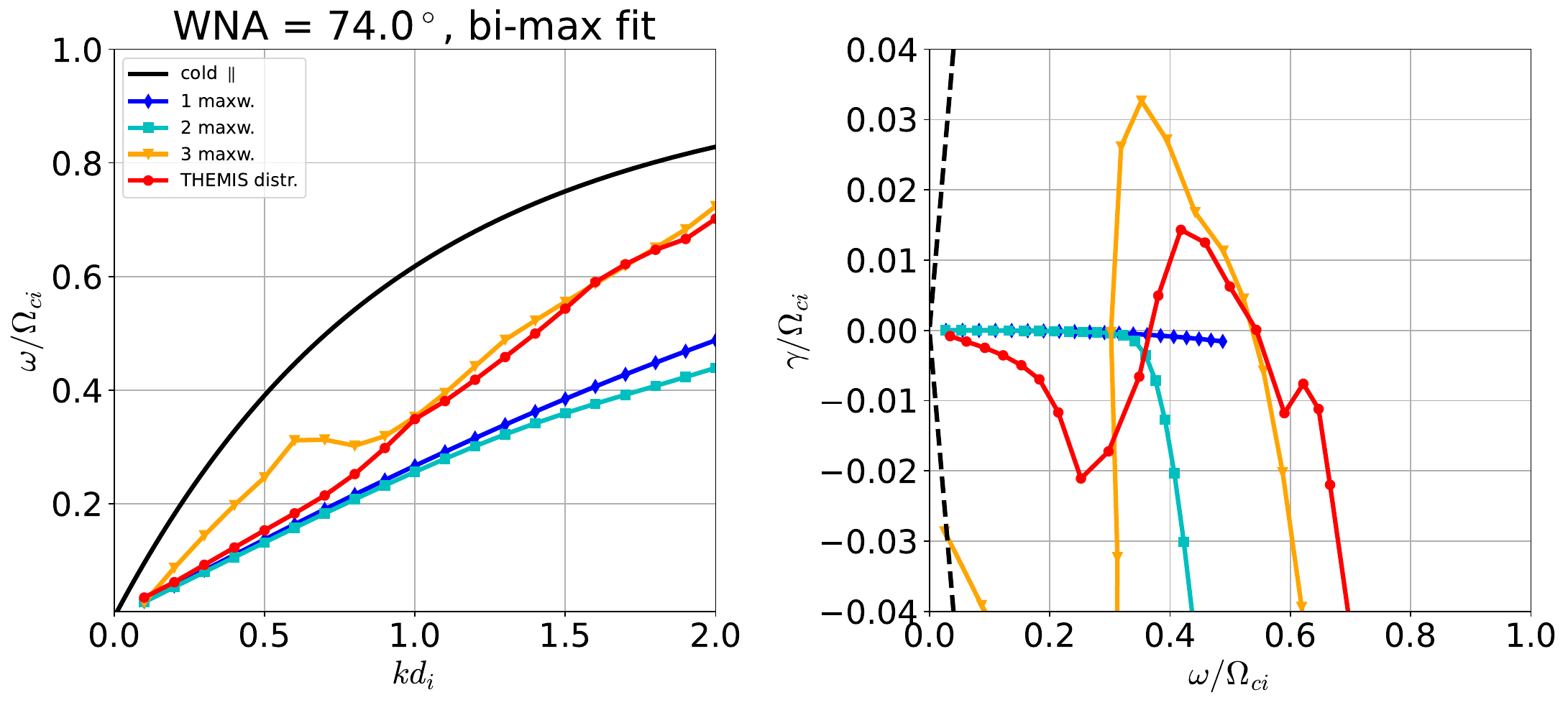} \caption{Solutions of the wave dispersion (left) and wave growth rate (right) for EMIC waves with $74^\circ$ wave normal angle: results for the observed ion distribution are shown in red, results for the three bi-maxwellian distributions from Figure \ref{fig:2020-02-17bi-max_fit} are shown in blue (cold population only), cyan (cold and thermal populations), and orange (cold, thermal, and hot populations), respectively. Left panel shows wave dispersion relation: $\Omega_{ci}$ is the proton cyclotron frequency, $d_i= c/\omega_{pi}$ is proton inertial length. Black line shows the analytical solution for field-aligned EMIC waves in cold proton-electron plasma with background parameters from observations during the event \cite{bookStix62}. Black dashed traces in the right panel show the absolute value of the growth rate, $|\gamma|=\omega$.}
    \label{fig:disp_bimax_0}
\end{figure}

To reveal the ion distribution that is responsible for the generation of oblique EMIC waves, we decompose the observed $(v_\parallel, v_\perp)$ 
 distribution into three ion populations, each fitted by a bi-maxwellian distribution. Figure \ref{fig:2020-02-17bi-max_fit} shows (a) the observed distribution, (b) the cold ion population with $\beta\sim 10^{-2}$ (this population does not contribute to EMIC wave generation, but is needed to realistically match the total ion density with observations), (c) sum of cold and thermal, with $\beta_\perp/\beta_\parallel \sim 0.45$, ion populations (the latter one is parallel anisotropic and used to represent the $<1$keV field-aligned ion streams), (d) sum of cold, thermal, and hot, with $\beta_\perp/\beta_\parallel \sim 3.8$, ion populations (the latter one is transversely anisotropic and provides free energy for EMIC generation). 
 Comparison of panels (a) and (d) in Fig. \ref{fig:2020-02-17bi-max_fit} shows that this three-component fitting agrees well with the observed ion distribution. Using this fitted distribution, we then evaluate the contribution of each ion population to the EMIC wave dispersion and generation. Figure \ref{fig:disp_bimax_0} shows the dispersion relations of EMIC waves with $74^\circ$ propagation angle from the observed ion distribution and from the fitted distributions in Figs. \ref{fig:2020-02-17bi-max_fit} (b-d). For comparison, we also show the cold plasma dispersion for field-aligned waves \cite{bookStix62}. Solutions of the wave dispersion for cold-only (blue trace) and cold $+$ thermal (cyan trace) populations are quite similar, apart from the stronger damping  in the latter. Introducing the hot transversely anisotropic population (orange trace) results in positive growth rate, which is larger than the growth rate from the observed ion distribution, but occupies a similar frequency range. The main role of the thermal, field-aligned anisotropic population is to provide strong cyclotron damping of low wave normal angle waves: note that the energy of the first cyclotron resonance decreases with wave normal angle decrease and reaches $1$keV (around the energy of the thermal population) for field-aligned waves. A secondary, yet important, role of the thermal population is in reducing the total ion anisotropy: although the anisotropy of the entire ion distribution, $\beta_\perp/\beta_\parallel=1.8$, does not exceed much the threshold for field-aligned EMIC generation \cite{Yue19:emic}, the anisotropy of the hot population is sufficiently large ($\beta_\perp/\beta_\parallel\approx 4$) to produce very oblique waves \cite<see the analogical mechanism of oblique whistler-mode wave generation by the highly anisotropic electron component in>{Gary11:pop}

\begin{figure}
    \centering
    \includegraphics[width=\textwidth]{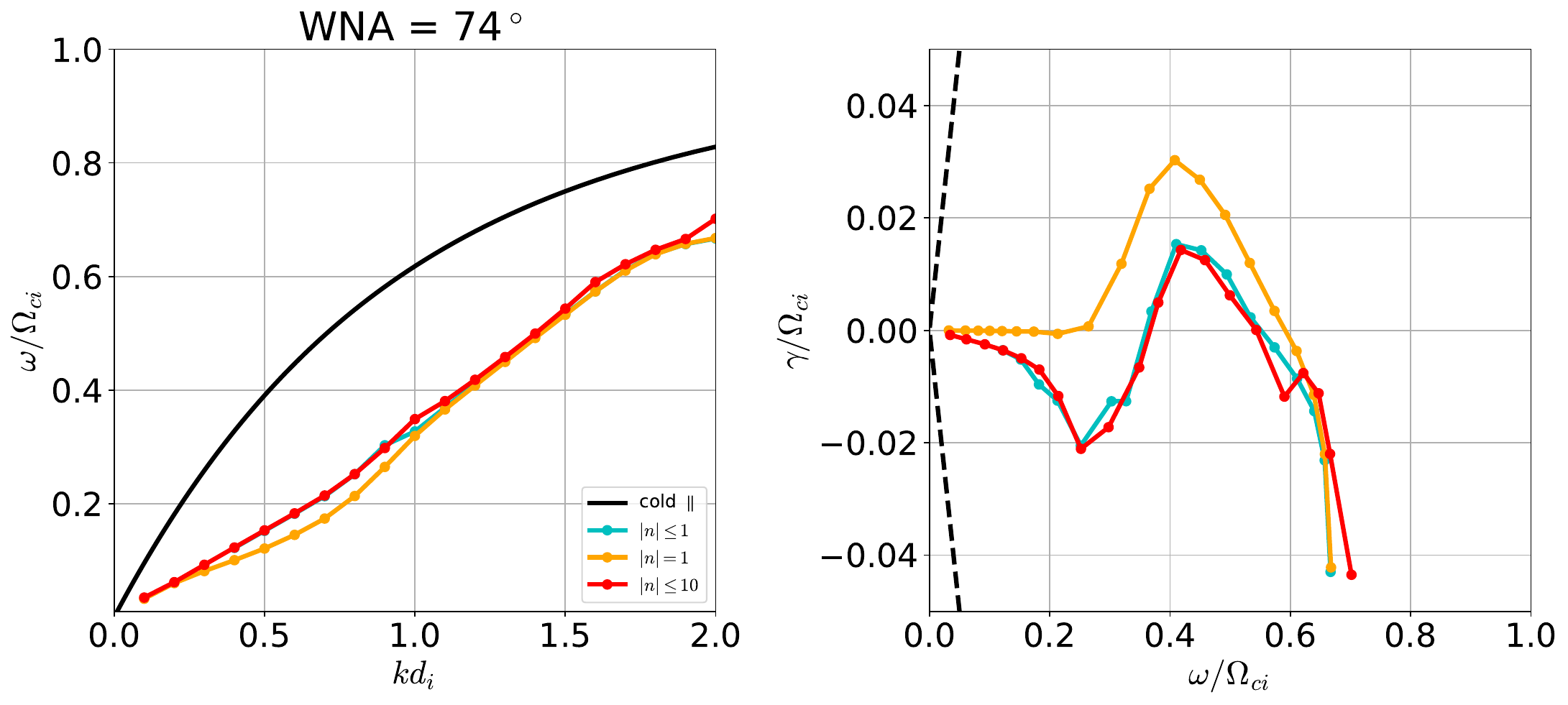} 
    \caption{Solutions of the wave dispersion (left) and wave growth rate (right) for $74^\circ$ wave normal angle, while keeping different orders of resonances in the dielectric tensor calculation. The result from the observed distribution for all resonances (in red, same as the red trace in Figure \ref{fig:disp_bimax_0}) includes up to $|n|=10$ resonances.}% with root finding error $10^{-3}$.}
    \label{fig:disp_res_number}
\end{figure}

We also investigate the contribution of different resonances to the oblique EMIC wave dispersion and growth rate. Figure \ref{fig:disp_res_number} shows the result for the first cyclotron resonance only, $n=1$ (orange), first cyclotron and Landau resonances $|n|\leq 1$ (cyan), and for all resonances up to $|n|=10$ (red). There is barely any difference between wave dispersions for these three cases, i.e., the wave dispersion is mainly provided by the ion population contributing to the first cyclotron resonance. The comparison of wave growth rates shows that the same ion population is responsible for wave generation, whereas Landau damping reduces the magnitude of the growth rate, especially at small wave frequencies. 

\begin{figure}
\begin{center}
\begin{tabular}{ c c }
 \includegraphics[width=0.5\textwidth]{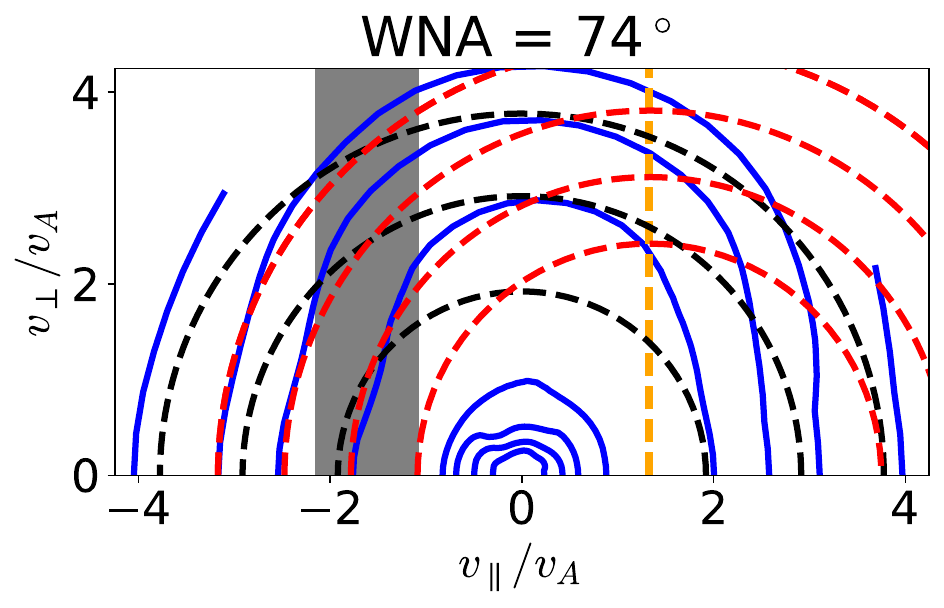} & \includegraphics[width=0.5\textwidth]{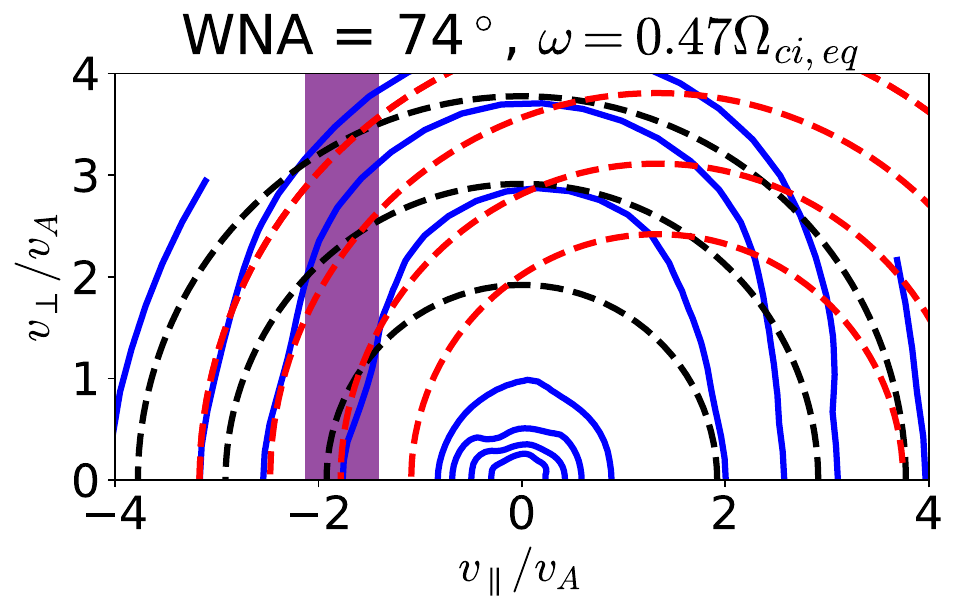} 
\end{tabular}
\end{center}
    \caption{Contours of constant phase space density from Figure \ref{fig:2020-02-17ion_distr} (blue), with resonant velocities for the first cyclotron resonance (shaded region), and resonant curves corresponding to contours of constant ion energy in the wave rest frame (red). Orange vertical line marks the Landau resonant energy. Left panel shows the range of resonant velocities at the equator for the frequency range with positive growth rate (grey shaded area), right panel shows the range of resonant velocities for a fixed equatorial frequency $0.47\Omega_{ci}$ and a latitudinal range of $0^\circ \leq\lambda\leq 25^\circ$ (purple shaded region).
    }
    \label{fig:2020-02-17distr_res_vel}
\end{figure}

Figure \ref{fig:2020-02-17distr_res_vel} further confirms the principal role of the first cyclotron resonance in generating the observed oblique EMIC waves, by combining contours of constant phase phase density (from Fig.  \ref{fig:2020-02-17ion_distr}) and resonant conditions for two approximations. The grey shaded region in the left panel shows the cyclotron resonant velocities calculated from the dispersion relation for $74^\circ$ wave normal angle and wave frequencies of positive growth rate, $\omega/\Omega_{ci} \in[0.38, 0.55]$. Comparison of constant phase phase density contours (blue traces) and resonance curves (contours of constant energy in the wave rest frame, shown in red) demonstrates that within the grey shaded region the gradient along the resonance curves (seen from the comparison of constant energy curves, shown in dashed black, and phase space density contours, shown in blue) corresponds to an increase in the phase space density, which will drive EMIC wave growth \cite{bookLyons&Williams}. Similarly, the same conclusion can be drawn for the resonant velocity range for monochromatic waves, with a frequency corresponding to the peak wave intensity in observations and variations of the resonance condition along magnetic latitudes (right panel): the positive gradient of the ion phase space density along resonance curves within the purple shaded region will amplify the waves as they propagate away from the generation region near the equator.

\subsection{Comparison of ion distributions during field-aligned and oblique EMIC events}

\begin{figure}
   \centering
   \includegraphics[width=\textwidth]{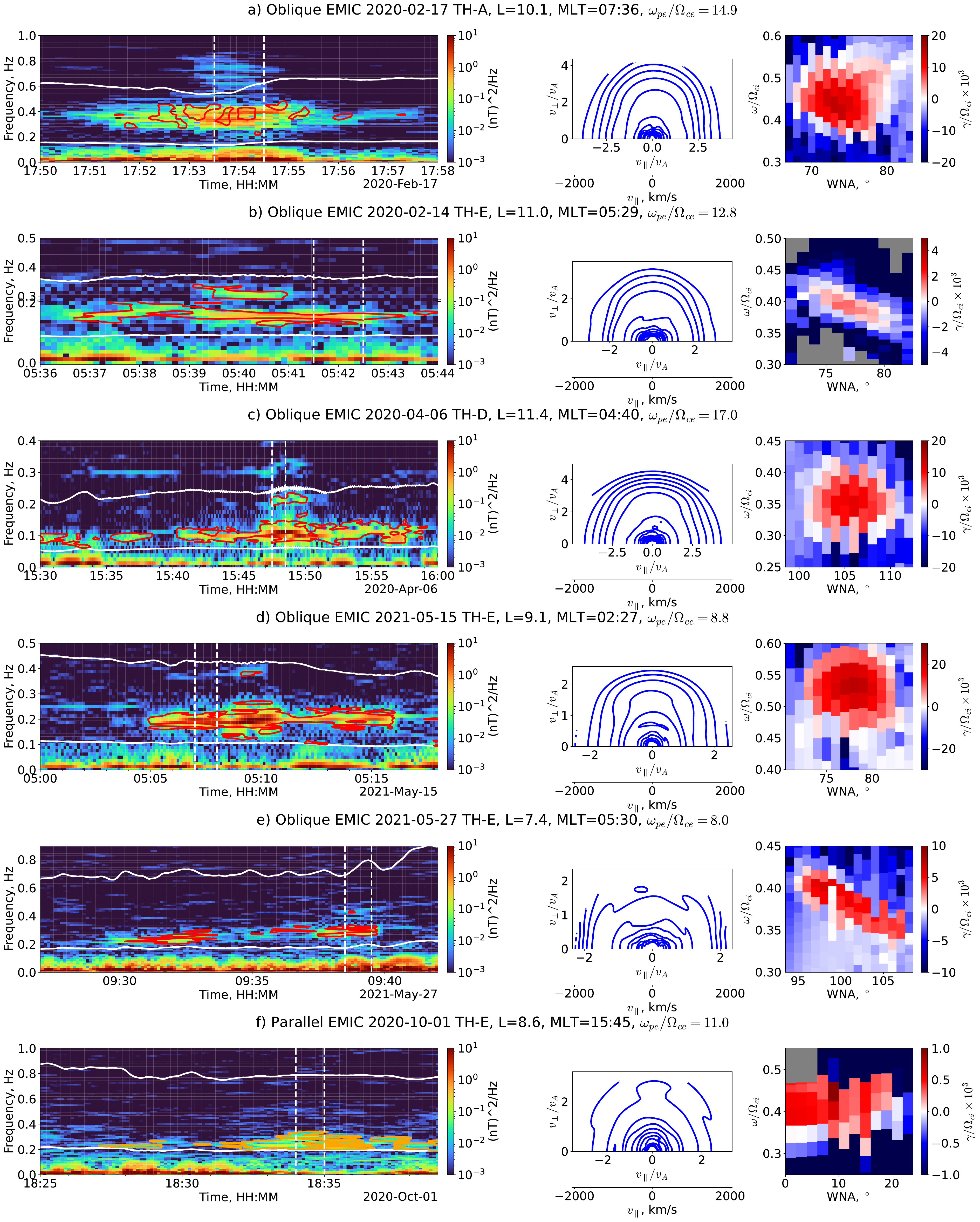} \caption{Wave magnetic field power spectrum (left column), ion distribution (center column), and wave growth rate (right column) for oblique (a-e) and field-aligned (f) EMIC wave events. In the wave power spectra: for oblique EMIC waves, red contours show the domain where wave normal angle $>60^\circ$, whereas for field-aligned EMIC wave event orange contours show the domain where wave normal angle $<20^\circ$; the two horizontal white lines mark hydrogen and helium ion gyrofrequencies from top to bottom. During the February 14, 2020 event (b), part of the wave spectrum has been cut to remove the spin tones that obscure the waves. }
   \label{fig:all_events}
\end{figure}

Figure \ref{fig:all_events}(b-e) shows four more examples of THEMIS observed very oblique EMIC waves in the dawn flank. All these events share similar properties of the event discussed in Section \ref{sec:event}:
\begin{itemize}
  \item Waves are proton band EMIC waves with clear maximum of the wave intensity around $[0.3,0.5]$ of local proton cyclotron frequency and with wave normal angles typically exceeding $60^\circ$ (left panels). 
  \item Ion distribution functions (middle panels) observed around the oblique EMIC wave burst include a very anisotropic hot (a few keVs) ion population ($\beta_\perp/\beta_\parallel>3$) and a field-aligned anisotropic thermal ($\leq 1$keV) ion population. The latter can lead to strong cyclotron damping of field-aligned EMIC waves. 
  \item Combining the measured ion distribution and the linear dispersion solver, LEOPARD, we show positive growth rates for very oblique EMIC waves (right panels), as well as damping for the field-aligned waves (not shown). Therefore, the most important condition for very oblique EMIC wave generation is likely the combination of transversely anisotropic hot ions (providing wave growth via cyclotron resonance) and field-aligned thermal ions (providing cyclotron damping of small wave normal angle waves).
\end{itemize}

To verify the importance of the thermal field-aligned population for very oblique EMIC wave growth, we further selected a typical field-aligned EMIC wave event observed by THEMIS in the dusk flank. As shown in Figure \ref{fig:all_events}(f), in the absence of field-aligned thermal ions, the field-aligned waves can be generated via cyclotron resonance with transversely anisotropic hot ions. Moreover, in the absence of the field-aligned population, the anisotropy of the entire ion distribution is mainly determined by the anisotropy of hot ions, which does not need to be as large as in events with oblique waves in order to generate field-aligned waves. As a result, in the event from Fig. \ref{fig:all_events}(f) the hot ion anisotropy, $\beta_\perp/\beta_\parallel=1.3$, is sufficient to generate field-aligned EMIC waves, but insufficient to generate oblique waves (the growth rate of oblique waves is negative for this event; not shown).

\section{Discussion and conclusions}\label{sec:discussion}
In this study we analyze several events with THEMIS observations of very oblique EMIC waves around the equator. During these events, the observed ion distributions consist of a highly transversely anisotropic ($\beta_\perp/\beta_\parallel\sim 4$, $>2$keV) hot ion population and a field-aligned anisotropic thermal ion population ($\beta_\perp/\beta_\parallel\sim 0.3$, $<1$keV). It is this field-aligned thermal ion population that prohibits the generation of field-aligned EMIC waves: the cyclotron resonant energy of waves with small wave normal angles is $\sim 1$keV, where the transverse anisotropy is insufficient to drive these waves. Therefore, such thermal ($<1$ keV) field-aligned ion population play a key role in producing very oblique EMIC waves, which are often observed on the dawn flank. The energy range and field-aligned anisotropy of this population suggest that these are likely ionospheric outflow ions \cite<see statistics of ion pitch-angle distributions in>[]{Artemyev18:jgr:RBSP&THEMIS,Yue17:ions}. Overlapping of such outflow, probably enhanced at the dawn flank due to strong plasma sheet electron precipitation driven by whistler-mode waves \cite{Thorne10:Nature,Ni16:ssr}, and the hot ion population, likely drifted from the dusk flank after being injected from the plasmasheet \cite<e.g.,>{Birn97:ion,Gabrielse14,Ukhorskiy18:DF}, creates favorable conditions for the generation of very oblique EMIC waves. This further implies that very oblique waves are likely a result of magnetosphere-ionosphere coupling, in contrast to the more typical field-aligned waves generated in the dusk flank due to plasmasheet injections or on the day side due to magnetosphere compression by the solar wind \cite{Yue19:emic, Jun19:emic,Jun21:emic}.

    \begin{figure}
        \centering
        \includegraphics[width = 0.8\textwidth]{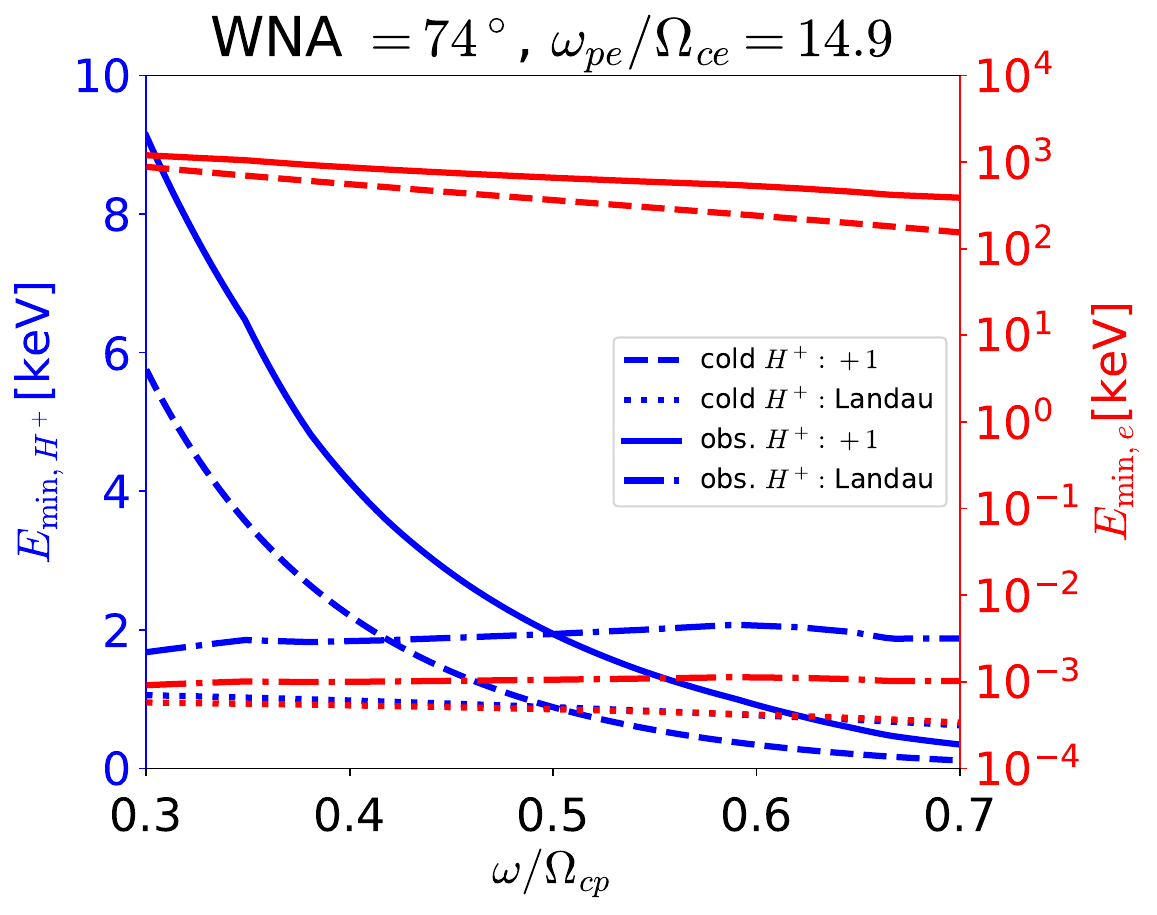}
        \caption{Minimum resonant energies for protons (blue, linear scale) and electrons (red, logarithmic scale) in Landau resonance and first cyclotron resonance ($"+1"$ for protons, $"-1"$ for electrons). Dashed and dotted lines are calculated using generic bi-maxwellian distributions (with $\beta= 10^{-2}$), whereas solid and dot-dashed lines are based on dispersion results for the first event (Figure \ref{fig:all_events} a).}
        \label{fig:energies}
    \end{figure}   

In addition to the possible resonant interactions between oblique EMIC waves and magnetospheric particles, as discussed in the introduction, a new resonant mechanism has been proposed recently that makes oblique EMIC waves potentially more important. It has been shown by \citeA{Hanzelka23:emic} that very oblique and sufficiently intense EMIC waves may resonate with energetic electrons via the so-called fractional (or subharmonic) resonances \cite{Lewak&Chen69,Smirnov&FrankKamenestkii68}. In contrast to cyclotron resonances with integer resonant numbers, the factional resonance is a purely nonlinear effect providing electron scattering in resonances with fractional numbers \cite{Terasawa&Matsukiyo12}, which reduces the electron energy in resonance with EMIC waves to sub-relativistic values \cite{Hanzelka23:emic}. THEMIS observations of very oblique EMIC waves and the proposed formation mechanism of these waves suggest that sub-relativistic electron precipitation on the dawn side \cite<see statistics of such precipitation in>[]{Tsai23} may be partly driven by EMIC waves, not exclusively by whistler-mode waves.

Figure \ref{fig:energies} shows the energies of electrons and ions in resonance with very oblique EMIC waves (we use a wave normal angle of $74^\circ$ and plot resonant energies as a function of wave frequency for typical plasma frequencies as from Figs. \ref{fig:all_events}). Note that the electron resonant energies for fractional resonances fall between the regions bounded by the first cyclotron resonance, $n=-1$, and Landau resonance, $n=0$ \cite{Hanzelka23:emic}. Comparing the results for the cold plasma dispersion ($\beta=10^{-2}$) and for the observed hot plasma with $\beta\approx 1.7$, we can see a higher resonant energy for the hot plasma case due to a decrease of the wave number at a fixed frequency \cite{Silin11}. Cyclotron and Landau resonant energies for protons are within $[1,10]$keV for $\omega/\Omega_{cp}<0.5$, where most of wave power is observed (see Fig. \ref{fig:all_events} (a)). This energy range allows oblique EMIC waves to heat thermal ($\sim 1$keV) ions and scatter ring current ($\sim 10$keV) ions into the loss cone (note that these resonant energies are calculated at the equator, which will increase in the off-equatorial region). The range of resonant energies for electrons, on the other hand, is much wider: from $\sim1$eV in Landau resonance to $1$MeV in cyclotron resonance. The fractional resonances with a resonance number $\in[0,1]$ will fill this gap and allow EMIC waves to also scatter hundreds of keV electrons. This is likely the most interesting and potentially important implication of very oblique waves.

To conclude, using THEMIS observations in the inner magnetosphere, we have investigated the generation mechanism of very oblique EMIC waves. Six typical events of such waves are observed at the dawn flank, in contrast to the more common field-aligned waves observed at the dusk and noon sectors. Very oblique EMIC waves are usually accompanied by ion distributions consisting of two main populations (except for the cold plasma population contributing to the total density). 1) A thermal ion population, at $\leq 1$ keV, with the field-line anisotropy providing the cyclotron damping of field-aligned EMIC waves and potentially reducing the Landau damping of oblique EMIC waves. This population shares the same properties of ionospheric outflow as reported previously for the inner magnetosphere \cite<e.g.,>{Yue17:ions}. 2) The hot ion population, at $>5$keV, with a strong transverse anisotropy ($\beta_\perp/\beta_\parallel\sim 4$) providing the cyclotron resonant growth of oblique EMIC waves with wave normal angles exceeding $60^\circ$. These observations underline the importance of magnetosphere-ionosphere coupling in producing very oblique EMIC waves. 

\acknowledgments
We acknowledge the support of NASA contract NAS5-02099 for the use of data from the THEMIS Mission, specifically K. H.Glassmeier, U. Auster and W. Baumjohann for the use of FGM data (provided under the lead of the Technical University of Braunschweig and with financial support through the German Ministry for Economy and Technology and the German Center for Aviation and Space (DLR) under contract 50 OC 0302).
Work of D.S.T., X.-J. Z., and A.V.A. are supported by NSF grant \#2329897, and NASA grants \#80NSSC20K1270, \#80NSSC23K0403, \#80NSSC23K0108 .

\section*{Open Research} \noindent 
THEMIS data is available at http://themis.ssl.berkeley.edu. Data access and processing was performed using SPEDAS V4.1 and its Python-based implementation, see \citeA{Angelopoulos19} and \cite{grimes2022space}.

%\bibliography{full,addon}

\begin{thebibliography}{}

\bibitem [\protect \citeauthoryear {%
{Angelopoulos}%
}{%
{Angelopoulos}%
}{%
{\protect \APACyear {2008}}%
}]{%
Angelopoulos08:ssr}
\APACinsertmetastar {%
Angelopoulos08:ssr}%
\begin{APACrefauthors}%
{Angelopoulos}, V.%
\end{APACrefauthors}%
\unskip\
\newblock
\APACrefYearMonthDay{2008}{{\APACmonth{12}}}{}.
\newblock
{\BBOQ}\APACrefatitle {{The THEMIS Mission}} {{The THEMIS Mission}}.{\BBCQ}
\newblock
\APACjournalVolNumPages{\ssr}{141}{}{5-34}.
\newblock
\begin{APACrefDOI} \doi{10.1007/s11214-008-9336-1} \end{APACrefDOI}
\PrintBackRefs{\CurrentBib}

\bibitem [\protect \citeauthoryear {%
{Angelopoulos}%
\ \protect \BOthers {.}}{%
{Angelopoulos}%
\ \protect \BOthers {.}}{%
{\protect \APACyear {2019}}%
}]{%
Angelopoulos19}
\APACinsertmetastar {%
Angelopoulos19}%
\begin{APACrefauthors}%
{Angelopoulos}, V.%
, {Cruce}, P.%
, {Drozdov}, A.%
, {Grimes}, E\BPBI W.%
, {Hatzigeorgiu}, N.%
, {King}, D\BPBI A.%
\BDBL {}{Schroeder}, P.%
\end{APACrefauthors}%
\unskip\
\newblock
\APACrefYearMonthDay{2019}{{\APACmonth{01}}}{}.
\newblock
{\BBOQ}\APACrefatitle {{The Space Physics Environment Data Analysis System (SPEDAS)}} {{The Space Physics Environment Data Analysis System (SPEDAS)}}.{\BBCQ}
\newblock
\APACjournalVolNumPages{\ssr}{215}{}{9}.
\newblock
\begin{APACrefDOI} \doi{10.1007/s11214-018-0576-4} \end{APACrefDOI}
\PrintBackRefs{\CurrentBib}

\bibitem [\protect \citeauthoryear {%
{Artemyev}%
\ \BBA {} {Mourenas}%
}{%
{Artemyev}%
\ \BBA {} {Mourenas}%
}{%
{\protect \APACyear {2020}}%
}]{%
Artemyev&Mourenas20:jgr}
\APACinsertmetastar {%
Artemyev&Mourenas20:jgr}%
\begin{APACrefauthors}%
{Artemyev}, A\BPBI V.%
\BCBT {}\ \BBA {} {Mourenas}, D.%
\end{APACrefauthors}%
\unskip\
\newblock
\APACrefYearMonthDay{2020}{{\APACmonth{03}}}{}.
\newblock
{\BBOQ}\APACrefatitle {{On Whistler Mode Wave Relation to Electron Field-Aligned Plateau Populations}} {{On Whistler Mode Wave Relation to Electron Field-Aligned Plateau Populations}}.{\BBCQ}
\newblock
\APACjournalVolNumPages{Journal of Geophysical Research (Space Physics)}{125}{3}{e27735}.
\newblock
\begin{APACrefDOI} \doi{10.1029/2019JA027735} \end{APACrefDOI}
\PrintBackRefs{\CurrentBib}

\bibitem [\protect \citeauthoryear {%
{Artemyev}%
\ \protect \BOthers {.}}{%
{Artemyev}%
\ \protect \BOthers {.}}{%
{\protect \APACyear {2018}}%
}]{%
Artemyev18:jgr:RBSP&THEMIS}
\APACinsertmetastar {%
Artemyev18:jgr:RBSP&THEMIS}%
\begin{APACrefauthors}%
{Artemyev}, A\BPBI V.%
, {Zhang}, X\BPBI J.%
, {Angelopoulos}, V.%
, {Runov}, A.%
, {Spence}, H\BPBI E.%
\BCBL {}\ \BBA {} {Larsen}, B\BPBI A.%
\end{APACrefauthors}%
\unskip\
\newblock
\APACrefYearMonthDay{2018}{Jul}{}.
\newblock
{\BBOQ}\APACrefatitle {{Plasma Anisotropies and Currents in the Near-Earth Plasma Sheet and Inner Magnetosphere}} {{Plasma Anisotropies and Currents in the Near-Earth Plasma Sheet and Inner Magnetosphere}}.{\BBCQ}
\newblock
\APACjournalVolNumPages{Journal of Geophysical Research (Space Physics)}{123}{7}{5625-5639}.
\newblock
\begin{APACrefDOI} \doi{10.1029/2018JA025232} \end{APACrefDOI}
\PrintBackRefs{\CurrentBib}

\bibitem [\protect \citeauthoryear {%
{Astfalk}%
\ \BBA {} {Jenko}%
}{%
{Astfalk}%
\ \BBA {} {Jenko}%
}{%
{\protect \APACyear {2017}}%
}]{%
Astfalk&Jenko17}
\APACinsertmetastar {%
Astfalk&Jenko17}%
\begin{APACrefauthors}%
{Astfalk}, P.%
\BCBT {}\ \BBA {} {Jenko}, F.%
\end{APACrefauthors}%
\unskip\
\newblock
\APACrefYearMonthDay{2017}{{\APACmonth{01}}}{}.
\newblock
{\BBOQ}\APACrefatitle {{LEOPARD: A grid-based dispersion relation solver for arbitrary gyrotropic distributions}} {{LEOPARD: A grid-based dispersion relation solver for arbitrary gyrotropic distributions}}.{\BBCQ}
\newblock
\APACjournalVolNumPages{Journal of Geophysical Research (Space Physics)}{122}{1}{89-101}.
\newblock
\begin{APACrefDOI} \doi{10.1002/2016JA023522} \end{APACrefDOI}
\PrintBackRefs{\CurrentBib}

\bibitem [\protect \citeauthoryear {%
{Auster}%
\ \protect \BOthers {.}}{%
{Auster}%
\ \protect \BOthers {.}}{%
{\protect \APACyear {2008}}%
}]{%
Auster08:THEMIS}
\APACinsertmetastar {%
Auster08:THEMIS}%
\begin{APACrefauthors}%
{Auster}, H\BPBI U.%
, {Glassmeier}, K\BPBI H.%
, {Magnes}, W.%
, {Aydogar}, O.%
, {Baumjohann}, W.%
, {Constantinescu}, D.%
\BDBL {}{Wiedemann}, M.%
\end{APACrefauthors}%
\unskip\
\newblock
\APACrefYearMonthDay{2008}{{\APACmonth{12}}}{}.
\newblock
{\BBOQ}\APACrefatitle {{The THEMIS Fluxgate Magnetometer}} {{The THEMIS Fluxgate Magnetometer}}.{\BBCQ}
\newblock
\APACjournalVolNumPages{\ssr}{141}{}{235-264}.
\newblock
\begin{APACrefDOI} \doi{10.1007/s11214-008-9365-9} \end{APACrefDOI}
\PrintBackRefs{\CurrentBib}

\bibitem [\protect \citeauthoryear {%
{Birn}%
\ \protect \BOthers {.}}{%
{Birn}%
\ \protect \BOthers {.}}{%
{\protect \APACyear {1997}}%
}]{%
Birn97:ion}
\APACinsertmetastar {%
Birn97:ion}%
\begin{APACrefauthors}%
{Birn}, J.%
, {Thomsen}, M\BPBI F.%
, {Borovsky}, J\BPBI E.%
, {Reeves}, G\BPBI D.%
, {McComas}, D\BPBI J.%
, {Belian}, R\BPBI D.%
\BCBL {}\ \BBA {} {Hesse}, M.%
\end{APACrefauthors}%
\unskip\
\newblock
\APACrefYearMonthDay{1997}{{\APACmonth{02}}}{}.
\newblock
{\BBOQ}\APACrefatitle {{Substorm ion injections: Geosynchronous observations and test particle orbits in three-dimensional dynamic MHD fields}} {{Substorm ion injections: Geosynchronous observations and test particle orbits in three-dimensional dynamic MHD fields}}.{\BBCQ}
\newblock
\APACjournalVolNumPages{\jgr}{102}{}{2325-2342}.
\newblock
\begin{APACrefDOI} \doi{10.1029/96JA03032} \end{APACrefDOI}
\PrintBackRefs{\CurrentBib}

\bibitem [\protect \citeauthoryear {%
{Blum}%
\ \protect \BOthers {.}}{%
{Blum}%
\ \protect \BOthers {.}}{%
{\protect \APACyear {2019}}%
}]{%
Blum19}
\APACinsertmetastar {%
Blum19}%
\begin{APACrefauthors}%
{Blum}, L\BPBI W.%
, {Artemyev}, A.%
, {Agapitov}, O.%
, {Mourenas}, D.%
, {Boardsen}, S.%
\BCBL {}\ \BBA {} {Schiller}, Q.%
\end{APACrefauthors}%
\unskip\
\newblock
\APACrefYearMonthDay{2019}{Apr}{}.
\newblock
{\BBOQ}\APACrefatitle {{EMIC Wave-Driven Bounce Resonance Scattering of Energetic Electrons in the Inner Magnetosphere}} {{EMIC Wave-Driven Bounce Resonance Scattering of Energetic Electrons in the Inner Magnetosphere}}.{\BBCQ}
\newblock
\APACjournalVolNumPages{Journal of Geophysical Research (Space Physics)}{124}{4}{2484-2496}.
\newblock
\begin{APACrefDOI} \doi{10.1029/2018JA026427} \end{APACrefDOI}
\PrintBackRefs{\CurrentBib}

\bibitem [\protect \citeauthoryear {%
{Bonnell}%
\ \protect \BOthers {.}}{%
{Bonnell}%
\ \protect \BOthers {.}}{%
{\protect \APACyear {2008}}%
}]{%
Bonnell08}
\APACinsertmetastar {%
Bonnell08}%
\begin{APACrefauthors}%
{Bonnell}, J\BPBI W.%
, {Mozer}, F\BPBI S.%
, {Delory}, G\BPBI T.%
, {Hull}, A\BPBI J.%
, {Ergun}, R\BPBI E.%
, {Cully}, C\BPBI M.%
\BDBL {}{Harvey}, P\BPBI R.%
\end{APACrefauthors}%
\unskip\
\newblock
\APACrefYearMonthDay{2008}{{\APACmonth{12}}}{}.
\newblock
{\BBOQ}\APACrefatitle {{The Electric Field Instrument (EFI) for THEMIS}} {{The Electric Field Instrument (EFI) for THEMIS}}.{\BBCQ}
\newblock
\APACjournalVolNumPages{\ssr}{141}{}{303-341}.
\newblock
\begin{APACrefDOI} \doi{10.1007/s11214-008-9469-2} \end{APACrefDOI}
\PrintBackRefs{\CurrentBib}

\bibitem [\protect \citeauthoryear {%
{Chen}%
, {Thorne}%
\BCBL {}\ \BBA {} {Bortnik}%
}{%
{Chen}%
\ \protect \BOthers {.}}{%
{\protect \APACyear {2011}}%
}]{%
Chen11:emic}
\APACinsertmetastar {%
Chen11:emic}%
\begin{APACrefauthors}%
{Chen}, L.%
, {Thorne}, R\BPBI M.%
\BCBL {}\ \BBA {} {Bortnik}, J.%
\end{APACrefauthors}%
\unskip\
\newblock
\APACrefYearMonthDay{2011}{{\APACmonth{08}}}{}.
\newblock
{\BBOQ}\APACrefatitle {{The controlling effect of ion temperature on EMIC wave excitation and scattering}} {{The controlling effect of ion temperature on EMIC wave excitation and scattering}}.{\BBCQ}
\newblock
\APACjournalVolNumPages{\grl}{38}{16}{L16109}.
\newblock
\begin{APACrefDOI} \doi{10.1029/2011GL048653} \end{APACrefDOI}
\PrintBackRefs{\CurrentBib}

\bibitem [\protect \citeauthoryear {%
{Chen}%
\ \protect \BOthers {.}}{%
{Chen}%
\ \protect \BOthers {.}}{%
{\protect \APACyear {2010}}%
}]{%
Chen10:emic}
\APACinsertmetastar {%
Chen10:emic}%
\begin{APACrefauthors}%
{Chen}, L.%
, {Thorne}, R\BPBI M.%
, {Jordanova}, V\BPBI K.%
, {Wang}, C\BHBI P.%
, {Gkioulidou}, M.%
, {Lyons}, L.%
\BCBL {}\ \BBA {} {Horne}, R\BPBI B.%
\end{APACrefauthors}%
\unskip\
\newblock
\APACrefYearMonthDay{2010}{Jul}{}.
\newblock
{\BBOQ}\APACrefatitle {{Global simulation of EMIC wave excitation during the 21 April 2001 storm from coupled RCM-RAM-HOTRAY modeling}} {{Global simulation of EMIC wave excitation during the 21 April 2001 storm from coupled RCM-RAM-HOTRAY modeling}}.{\BBCQ}
\newblock
\APACjournalVolNumPages{Journal of Geophysical Research (Space Physics)}{115}{A7}{A07209}.
\newblock
\begin{APACrefDOI} \doi{10.1029/2009JA015075} \end{APACrefDOI}
\PrintBackRefs{\CurrentBib}

\bibitem [\protect \citeauthoryear {%
{Cornwall}%
, {Coroniti}%
\BCBL {}\ \BBA {} {Thorne}%
}{%
{Cornwall}%
\ \protect \BOthers {.}}{%
{\protect \APACyear {1970}}%
}]{%
Cornwall70}
\APACinsertmetastar {%
Cornwall70}%
\begin{APACrefauthors}%
{Cornwall}, J\BPBI M.%
, {Coroniti}, F\BPBI V.%
\BCBL {}\ \BBA {} {Thorne}, R\BPBI M.%
\end{APACrefauthors}%
\unskip\
\newblock
\APACrefYearMonthDay{1970}{}{}.
\newblock
{\BBOQ}\APACrefatitle {{Turbulent loss of ring current protons}} {{Turbulent loss of ring current protons}}.{\BBCQ}
\newblock
\APACjournalVolNumPages{\jgr}{75}{}{4699}.
\newblock
\begin{APACrefDOI} \doi{10.1029/JA075i025p04699} \end{APACrefDOI}
\PrintBackRefs{\CurrentBib}

\bibitem [\protect \citeauthoryear {%
{Cornwall}%
, {Coroniti}%
\BCBL {}\ \BBA {} {Thorne}%
}{%
{Cornwall}%
\ \protect \BOthers {.}}{%
{\protect \APACyear {1971}}%
}]{%
Cornwall71:arc}
\APACinsertmetastar {%
Cornwall71:arc}%
\begin{APACrefauthors}%
{Cornwall}, J\BPBI M.%
, {Coroniti}, F\BPBI V.%
\BCBL {}\ \BBA {} {Thorne}, R\BPBI M.%
\end{APACrefauthors}%
\unskip\
\newblock
\APACrefYearMonthDay{1971}{{\APACmonth{01}}}{}.
\newblock
{\BBOQ}\APACrefatitle {{Unified theory of SAR arc formation at the plasmapause}} {{Unified theory of SAR arc formation at the plasmapause}}.{\BBCQ}
\newblock
\APACjournalVolNumPages{\jgr}{76}{19}{4428}.
\newblock
\begin{APACrefDOI} \doi{10.1029/JA076i019p04428} \end{APACrefDOI}
\PrintBackRefs{\CurrentBib}

\bibitem [\protect \citeauthoryear {%
{Cornwall}%
\ \BBA {} {Schulz}%
}{%
{Cornwall}%
\ \BBA {} {Schulz}%
}{%
{\protect \APACyear {1971}}%
}]{%
Cornwall&Schulz71}
\APACinsertmetastar {%
Cornwall&Schulz71}%
\begin{APACrefauthors}%
{Cornwall}, J\BPBI M.%
\BCBT {}\ \BBA {} {Schulz}, M.%
\end{APACrefauthors}%
\unskip\
\newblock
\APACrefYearMonthDay{1971}{{\APACmonth{11}}}{}.
\newblock
{\BBOQ}\APACrefatitle {{Electromagnetic ion-cyclotron instabilities in multicomponent magnetospheric plasmas}} {{Electromagnetic ion-cyclotron instabilities in multicomponent magnetospheric plasmas}}.{\BBCQ}
\newblock
\APACjournalVolNumPages{\jgr}{76}{31}{7791-7796}.
\newblock
\begin{APACrefDOI} \doi{10.1029/JA076i031p07791} \end{APACrefDOI}
\PrintBackRefs{\CurrentBib}

\bibitem [\protect \citeauthoryear {%
{de Soria-Santacruz}%
, {Spasojevic}%
\BCBL {}\ \BBA {} {Chen}%
}{%
{de Soria-Santacruz}%
\ \protect \BOthers {.}}{%
{\protect \APACyear {2013}}%
}]{%
SoriaSantacruz13}
\APACinsertmetastar {%
SoriaSantacruz13}%
\begin{APACrefauthors}%
{de Soria-Santacruz}, M.%
, {Spasojevic}, M.%
\BCBL {}\ \BBA {} {Chen}, L.%
\end{APACrefauthors}%
\unskip\
\newblock
\APACrefYearMonthDay{2013}{{\APACmonth{05}}}{}.
\newblock
{\BBOQ}\APACrefatitle {{EMIC waves growth and guiding in the presence of cold plasma density irregularities}} {{EMIC waves growth and guiding in the presence of cold plasma density irregularities}}.{\BBCQ}
\newblock
\APACjournalVolNumPages{\grl}{40}{10}{1940-1944}.
\newblock
\begin{APACrefDOI} \doi{10.1002/grl.50484} \end{APACrefDOI}
\PrintBackRefs{\CurrentBib}

\bibitem [\protect \citeauthoryear {%
{Fraser}%
\ \BBA {} {Nguyen}%
}{%
{Fraser}%
\ \BBA {} {Nguyen}%
}{%
{\protect \APACyear {2001}}%
}]{%
Fraser&Nguyen01}
\APACinsertmetastar {%
Fraser&Nguyen01}%
\begin{APACrefauthors}%
{Fraser}, B\BPBI J.%
\BCBT {}\ \BBA {} {Nguyen}, T\BPBI S.%
\end{APACrefauthors}%
\unskip\
\newblock
\APACrefYearMonthDay{2001}{{\APACmonth{07}}}{}.
\newblock
{\BBOQ}\APACrefatitle {{Is the plasmapause a preferred source region of electromagnetic ion cyclotron waves in the magnetosphere?}} {{Is the plasmapause a preferred source region of electromagnetic ion cyclotron waves in the magnetosphere?}}{\BBCQ}
\newblock
\APACjournalVolNumPages{Journal of Atmospheric and Solar-Terrestrial Physics}{63}{11}{1225-1247}.
\newblock
\begin{APACrefDOI} \doi{10.1016/S1364-6826(00)00225-X} \end{APACrefDOI}
\PrintBackRefs{\CurrentBib}

\bibitem [\protect \citeauthoryear {%
{Gabrielse}%
, {Angelopoulos}%
, {Runov}%
\BCBL {}\ \BBA {} {Turner}%
}{%
{Gabrielse}%
\ \protect \BOthers {.}}{%
{\protect \APACyear {2014}}%
}]{%
Gabrielse14}
\APACinsertmetastar {%
Gabrielse14}%
\begin{APACrefauthors}%
{Gabrielse}, C.%
, {Angelopoulos}, V.%
, {Runov}, A.%
\BCBL {}\ \BBA {} {Turner}, D\BPBI L.%
\end{APACrefauthors}%
\unskip\
\newblock
\APACrefYearMonthDay{2014}{{\APACmonth{04}}}{}.
\newblock
{\BBOQ}\APACrefatitle {{Statistical characteristics of particle injections throughout the equatorial magnetotail}} {{Statistical characteristics of particle injections throughout the equatorial magnetotail}}.{\BBCQ}
\newblock
\APACjournalVolNumPages{\jgr}{119}{}{2512-2535}.
\newblock
\begin{APACrefDOI} \doi{10.1002/2013JA019638} \end{APACrefDOI}
\PrintBackRefs{\CurrentBib}

\bibitem [\protect \citeauthoryear {%
{Gary}%
, {Liu}%
\BCBL {}\ \BBA {} {Winske}%
}{%
{Gary}%
\ \protect \BOthers {.}}{%
{\protect \APACyear {2011}}%
}]{%
Gary11:pop}
\APACinsertmetastar {%
Gary11:pop}%
\begin{APACrefauthors}%
{Gary}, S\BPBI P.%
, {Liu}, K.%
\BCBL {}\ \BBA {} {Winske}, D.%
\end{APACrefauthors}%
\unskip\
\newblock
\APACrefYearMonthDay{2011}{{\APACmonth{08}}}{}.
\newblock
{\BBOQ}\APACrefatitle {{Whistler anisotropy instability at low electron {\ensuremath{\beta}}: Particle-in-cell simulations}} {{Whistler anisotropy instability at low electron {\ensuremath{\beta}}: Particle-in-cell simulations}}.{\BBCQ}
\newblock
\APACjournalVolNumPages{Physics of Plasmas}{18}{8}{082902}.
\newblock
\begin{APACrefDOI} \doi{10.1063/1.3610378} \end{APACrefDOI}
\PrintBackRefs{\CurrentBib}

\bibitem [\protect \citeauthoryear {%
Grimes%
\ \protect \BOthers {.}}{%
Grimes%
\ \protect \BOthers {.}}{%
{\protect \APACyear {2022}}%
}]{%
grimes2022space}
\APACinsertmetastar {%
grimes2022space}%
\begin{APACrefauthors}%
Grimes, E\BPBI W.%
, Harter, B.%
, Hatzigeorgiu, N.%
, Drozdov, A.%
, Lewis, J\BPBI W.%
, Angelopoulos, V.%
\BDBL {}others%
\end{APACrefauthors}%
\unskip\
\newblock
\APACrefYearMonthDay{2022}{}{}.
\newblock
{\BBOQ}\APACrefatitle {The space physics environment data analysis system in Python} {The space physics environment data analysis system in python}.{\BBCQ}
\newblock
\APACjournalVolNumPages{Frontiers in Astronomy and Space Sciences}{9}{}{1020815}.
\PrintBackRefs{\CurrentBib}

\bibitem [\protect \citeauthoryear {%
{Hanzelka}%
, {Li}%
\BCBL {}\ \BBA {} {Ma}%
}{%
{Hanzelka}%
\ \protect \BOthers {.}}{%
{\protect \APACyear {2023}}%
{\protect \APACexlab {{\protect \BCnt {1}}}}}]{%
Hanzelka23}
\APACinsertmetastar {%
Hanzelka23}%
\begin{APACrefauthors}%
{Hanzelka}, M.%
, {Li}, W.%
\BCBL {}\ \BBA {} {Ma}, Q.%
\end{APACrefauthors}%
\unskip\
\newblock
\APACrefYearMonthDay{2023{\protect \BCnt {1}}}{{\APACmonth{04}}}{}.
\newblock
{\BBOQ}\APACrefatitle {{Parametric analysis of pitch angle scattering and losses of relativistic electrons by oblique EMIC waves}} {{Parametric analysis of pitch angle scattering and losses of relativistic electrons by oblique EMIC waves}}.{\BBCQ}
\newblock
\APACjournalVolNumPages{Frontiers in Astronomy and Space Sciences}{10}{}{1163515}.
\newblock
\begin{APACrefDOI} \doi{10.3389/fspas.2023.1163515} \end{APACrefDOI}
\PrintBackRefs{\CurrentBib}

\bibitem [\protect \citeauthoryear {%
{Hanzelka}%
, {Li}%
\BCBL {}\ \BBA {} {Ma}%
}{%
{Hanzelka}%
\ \protect \BOthers {.}}{%
{\protect \APACyear {2023}}%
{\protect \APACexlab {{\protect \BCnt {2}}}}}]{%
Hanzelka23:emic}
\APACinsertmetastar {%
Hanzelka23:emic}%
\begin{APACrefauthors}%
{Hanzelka}, M.%
, {Li}, W.%
\BCBL {}\ \BBA {} {Ma}, Q.%
\end{APACrefauthors}%
\unskip\
\newblock
\APACrefYearMonthDay{2023{\protect \BCnt {2}}}{{\APACmonth{04}}}{}.
\newblock
{\BBOQ}\APACrefatitle {{Parametric analysis of pitch angle scattering and losses of relativistic electrons by oblique EMIC waves}} {{Parametric analysis of pitch angle scattering and losses of relativistic electrons by oblique EMIC waves}}.{\BBCQ}
\newblock
\APACjournalVolNumPages{Frontiers in Astronomy and Space Sciences}{10}{}{1163515}.
\newblock
\begin{APACrefDOI} \doi{10.3389/fspas.2023.1163515} \end{APACrefDOI}
\PrintBackRefs{\CurrentBib}

\bibitem [\protect \citeauthoryear {%
{Jun}%
\ \protect \BOthers {.}}{%
{Jun}%
\ \protect \BOthers {.}}{%
{\protect \APACyear {2021}}%
}]{%
Jun21:emic}
\APACinsertmetastar {%
Jun21:emic}%
\begin{APACrefauthors}%
{Jun}, C\BHBI W.%
, {Miyoshi}, Y.%
, {Kurita}, S.%
, {Yue}, C.%
, {Bortnik}, J.%
, {Lyons}, L.%
\BDBL {}{Shinohara}, I.%
\end{APACrefauthors}%
\unskip\
\newblock
\APACrefYearMonthDay{2021}{{\APACmonth{06}}}{}.
\newblock
{\BBOQ}\APACrefatitle {{The Characteristics of EMIC Waves in the Magnetosphere Based on the Van Allen Probes and Arase Observations}} {{The Characteristics of EMIC Waves in the Magnetosphere Based on the Van Allen Probes and Arase Observations}}.{\BBCQ}
\newblock
\APACjournalVolNumPages{Journal of Geophysical Research (Space Physics)}{126}{6}{e29001}.
\newblock
\begin{APACrefDOI} \doi{10.1029/2020JA029001} \end{APACrefDOI}
\PrintBackRefs{\CurrentBib}

\bibitem [\protect \citeauthoryear {%
{Jun}%
\ \protect \BOthers {.}}{%
{Jun}%
\ \protect \BOthers {.}}{%
{\protect \APACyear {2019}}%
}]{%
Jun19:emic}
\APACinsertmetastar {%
Jun19:emic}%
\begin{APACrefauthors}%
{Jun}, C\BPBI W.%
, {Yue}, C.%
, {Bortnik}, J.%
, {Lyons}, L\BPBI R.%
, {Nishimura}, Y.%
\BCBL {}\ \BBA {} {Kletzing}, C.%
\end{APACrefauthors}%
\unskip\
\newblock
\APACrefYearMonthDay{2019}{Mar}{}.
\newblock
{\BBOQ}\APACrefatitle {{EMIC Wave Properties Associated With and Without Injections in The Inner Magnetosphere}} {{EMIC Wave Properties Associated With and Without Injections in The Inner Magnetosphere}}.{\BBCQ}
\newblock
\APACjournalVolNumPages{Journal of Geophysical Research (Space Physics)}{124}{3}{2029-2045}.
\newblock
\begin{APACrefDOI} \doi{10.1029/2018JA026279} \end{APACrefDOI}
\PrintBackRefs{\CurrentBib}

\bibitem [\protect \citeauthoryear {%
{Khazanov}%
\ \BBA {} {Gamayunov}%
}{%
{Khazanov}%
\ \BBA {} {Gamayunov}%
}{%
{\protect \APACyear {2007}}%
}]{%
Khazanov&Gamayunov07}
\APACinsertmetastar {%
Khazanov&Gamayunov07}%
\begin{APACrefauthors}%
{Khazanov}, G\BPBI V.%
\BCBT {}\ \BBA {} {Gamayunov}, K\BPBI V.%
\end{APACrefauthors}%
\unskip\
\newblock
\APACrefYearMonthDay{2007}{{\APACmonth{10}}}{}.
\newblock
{\BBOQ}\APACrefatitle {{Effect of electromagnetic ion cyclotron wave normal angle distribution on relativistic electron scattering in outer radiation belt}} {{Effect of electromagnetic ion cyclotron wave normal angle distribution on relativistic electron scattering in outer radiation belt}}.{\BBCQ}
\newblock
\APACjournalVolNumPages{Journal of Geophysical Research (Space Physics)}{112}{A10}{A10209}.
\newblock
\begin{APACrefDOI} \doi{10.1029/2007JA012282} \end{APACrefDOI}
\PrintBackRefs{\CurrentBib}

\bibitem [\protect \citeauthoryear {%
{Kim}%
\ \protect \BOthers {.}}{%
{Kim}%
\ \protect \BOthers {.}}{%
{\protect \APACyear {2021}}%
}]{%
Kim21:emic}
\APACinsertmetastar {%
Kim21:emic}%
\begin{APACrefauthors}%
{Kim}, H.%
, {Schiller}, Q.%
, {Engebretson}, M\BPBI J.%
, {Noh}, S.%
, {Kuzichev}, I.%
, {Lanzerotti}, L\BPBI J.%
\BDBL {}{Fromm}, T.%
\end{APACrefauthors}%
\unskip\
\newblock
\APACrefYearMonthDay{2021}{{\APACmonth{02}}}{}.
\newblock
{\BBOQ}\APACrefatitle {{Observations of Particle Loss due to Injection Associated Electromagnetic Ion Cyclotron Waves}} {{Observations of Particle Loss due to Injection Associated Electromagnetic Ion Cyclotron Waves}}.{\BBCQ}
\newblock
\APACjournalVolNumPages{Journal of Geophysical Research (Space Physics)}{126}{2}{e28503}.
\newblock
\begin{APACrefDOI} \doi{10.1029/2020JA028503} \end{APACrefDOI}
\PrintBackRefs{\CurrentBib}

\bibitem [\protect \citeauthoryear {%
{Kitamura}%
\ \protect \BOthers {.}}{%
{Kitamura}%
\ \protect \BOthers {.}}{%
{\protect \APACyear {2018}}%
}]{%
Kitamura18}
\APACinsertmetastar {%
Kitamura18}%
\begin{APACrefauthors}%
{Kitamura}, N.%
, {Kitahara}, M.%
, {Shoji}, M.%
, {Miyoshi}, Y.%
, {Hasegawa}, H.%
, {Nakamura}, S.%
\BDBL {}{Burch}, J\BPBI L.%
\end{APACrefauthors}%
\unskip\
\newblock
\APACrefYearMonthDay{2018}{{\APACmonth{09}}}{}.
\newblock
{\BBOQ}\APACrefatitle {{Direct measurements of two-way wave-particle energy transfer in a collisionless space plasma}} {{Direct measurements of two-way wave-particle energy transfer in a collisionless space plasma}}.{\BBCQ}
\newblock
\APACjournalVolNumPages{Science}{361}{6406}{1000-1003}.
\newblock
\begin{APACrefDOI} \doi{10.1126/science.aap8730} \end{APACrefDOI}
\PrintBackRefs{\CurrentBib}

\bibitem [\protect \citeauthoryear {%
{Lee}%
, {Shin}%
\BCBL {}\ \BBA {} {Choi}%
}{%
{Lee}%
\ \protect \BOthers {.}}{%
{\protect \APACyear {2018}}%
}]{%
Lee18:emic}
\APACinsertmetastar {%
Lee18:emic}%
\begin{APACrefauthors}%
{Lee}, D\BHBI Y.%
, {Shin}, D\BHBI K.%
\BCBL {}\ \BBA {} {Choi}, C\BHBI R.%
\end{APACrefauthors}%
\unskip\
\newblock
\APACrefYearMonthDay{2018}{{\APACmonth{06}}}{}.
\newblock
{\BBOQ}\APACrefatitle {{Effects of Oblique Wave Normal Angle and Noncircular Polarization of Electromagnetic Ion Cyclotron Waves on the Pitch Angle Scattering of Relativistic Electrons}} {{Effects of Oblique Wave Normal Angle and Noncircular Polarization of Electromagnetic Ion Cyclotron Waves on the Pitch Angle Scattering of Relativistic Electrons}}.{\BBCQ}
\newblock
\APACjournalVolNumPages{Journal of Geophysical Research (Space Physics)}{123}{6}{4556-4573}.
\newblock
\begin{APACrefDOI} \doi{10.1029/2018JA025342} \end{APACrefDOI}
\PrintBackRefs{\CurrentBib}

\bibitem [\protect \citeauthoryear {%
{Lewak}%
\ \BBA {} {Chen}%
}{%
{Lewak}%
\ \BBA {} {Chen}%
}{%
{\protect \APACyear {1969}}%
}]{%
Lewak&Chen69}
\APACinsertmetastar {%
Lewak&Chen69}%
\begin{APACrefauthors}%
{Lewak}, G\BPBI J.%
\BCBT {}\ \BBA {} {Chen}, C\BPBI S.%
\end{APACrefauthors}%
\unskip\
\newblock
\APACrefYearMonthDay{1969}{{\APACmonth{09}}}{}.
\newblock
{\BBOQ}\APACrefatitle {{Higher order resonances in a plasma}} {{Higher order resonances in a plasma}}.{\BBCQ}
\newblock
\APACjournalVolNumPages{Journal of Plasma Physics}{3}{3}{481-497}.
\newblock
\begin{APACrefDOI} \doi{10.1017/S0022377800004554} \end{APACrefDOI}
\PrintBackRefs{\CurrentBib}

\bibitem [\protect \citeauthoryear {%
{Li}%
\ \protect \BOthers {.}}{%
{Li}%
\ \protect \BOthers {.}}{%
{\protect \APACyear {2016}}%
}]{%
Li16}
\APACinsertmetastar {%
Li16}%
\begin{APACrefauthors}%
{Li}, W.%
, {Mourenas}, D.%
, {Artemyev}, A\BPBI V.%
, {Bortnik}, J.%
, {Thorne}, R\BPBI M.%
, {Kletzing}, C\BPBI A.%
\BDBL {}{Spence}, H\BPBI E.%
\end{APACrefauthors}%
\unskip\
\newblock
\APACrefYearMonthDay{2016}{{\APACmonth{09}}}{}.
\newblock
{\BBOQ}\APACrefatitle {{Unraveling the excitation mechanisms of highly oblique lower band chorus waves}} {{Unraveling the excitation mechanisms of highly oblique lower band chorus waves}}.{\BBCQ}
\newblock
\APACjournalVolNumPages{\grl}{43}{}{8867-8875}.
\newblock
\begin{APACrefDOI} \doi{10.1002/2016GL070386} \end{APACrefDOI}
\PrintBackRefs{\CurrentBib}

\bibitem [\protect \citeauthoryear {%
{Liu}%
, {Fraser}%
\BCBL {}\ \BBA {} {Menk}%
}{%
{Liu}%
\ \protect \BOthers {.}}{%
{\protect \APACyear {2013}}%
}]{%
Liu13:emic}
\APACinsertmetastar {%
Liu13:emic}%
\begin{APACrefauthors}%
{Liu}, Y\BPBI H.%
, {Fraser}, B\BPBI J.%
\BCBL {}\ \BBA {} {Menk}, F\BPBI W.%
\end{APACrefauthors}%
\unskip\
\newblock
\APACrefYearMonthDay{2013}{{\APACmonth{09}}}{}.
\newblock
{\BBOQ}\APACrefatitle {{EMIC waves observed by Cluster near the plasmapause}} {{EMIC waves observed by Cluster near the plasmapause}}.{\BBCQ}
\newblock
\APACjournalVolNumPages{Journal of Geophysical Research (Space Physics)}{118}{9}{5603-5615}.
\newblock
\begin{APACrefDOI} \doi{10.1002/jgra.50486} \end{APACrefDOI}
\PrintBackRefs{\CurrentBib}

\bibitem [\protect \citeauthoryear {%
{Lyons}%
\ \BBA {} {Williams}%
}{%
{Lyons}%
\ \BBA {} {Williams}%
}{%
{\protect \APACyear {1984}}%
}]{%
bookLyons&Williams}
\APACinsertmetastar {%
bookLyons&Williams}%
\begin{APACrefauthors}%
{Lyons}, L\BPBI R.%
\BCBT {}\ \BBA {} {Williams}, D\BPBI J.%
\end{APACrefauthors}%
\unskip\
\newblock
\APACrefYear{1984}.
\newblock
\APACrefbtitle {{Quantitative aspects of magnetospheric physics.}} {{Quantitative aspects of magnetospheric physics.}}\ ({Lyons, L.~R.~\& Williams, D.~J.}, \BED{}).
\PrintBackRefs{\CurrentBib}

\bibitem [\protect \citeauthoryear {%
{Ma}%
\ \protect \BOthers {.}}{%
{Ma}%
\ \protect \BOthers {.}}{%
{\protect \APACyear {2019}}%
}]{%
Ma19}
\APACinsertmetastar {%
Ma19}%
\begin{APACrefauthors}%
{Ma}, Q.%
, {Li}, W.%
, {Yue}, C.%
, {Thorne}, R\BPBI M.%
, {Bortnik}, J.%
, {Kletzing}, C\BPBI A.%
\BDBL {}{Spence}, H\BPBI E.%
\end{APACrefauthors}%
\unskip\
\newblock
\APACrefYearMonthDay{2019}{{\APACmonth{06}}}{}.
\newblock
{\BBOQ}\APACrefatitle {{Ion Heating by Electromagnetic Ion Cyclotron Waves and Magnetosonic Waves in the Earth's Inner Magnetosphere}} {{Ion Heating by Electromagnetic Ion Cyclotron Waves and Magnetosonic Waves in the Earth's Inner Magnetosphere}}.{\BBCQ}
\newblock
\APACjournalVolNumPages{\grl}{46}{12}{6258-6267}.
\newblock
\begin{APACrefDOI} \doi{10.1029/2019GL083513} \end{APACrefDOI}
\PrintBackRefs{\CurrentBib}

\bibitem [\protect \citeauthoryear {%
{McFadden}%
\ \protect \BOthers {.}}{%
{McFadden}%
\ \protect \BOthers {.}}{%
{\protect \APACyear {2008}}%
}]{%
McFadden08:THEMIS}
\APACinsertmetastar {%
McFadden08:THEMIS}%
\begin{APACrefauthors}%
{McFadden}, J\BPBI P.%
, {Carlson}, C\BPBI W.%
, {Larson}, D.%
, {Ludlam}, M.%
, {Abiad}, R.%
, {Elliott}, B.%
\BDBL {}{Angelopoulos}, V.%
\end{APACrefauthors}%
\unskip\
\newblock
\APACrefYearMonthDay{2008}{{\APACmonth{12}}}{}.
\newblock
{\BBOQ}\APACrefatitle {{The THEMIS ESA Plasma Instrument and In-flight Calibration}} {{The THEMIS ESA Plasma Instrument and In-flight Calibration}}.{\BBCQ}
\newblock
\APACjournalVolNumPages{\ssr}{141}{}{277-302}.
\newblock
\begin{APACrefDOI} \doi{10.1007/s11214-008-9440-2} \end{APACrefDOI}
\PrintBackRefs{\CurrentBib}

\bibitem [\protect \citeauthoryear {%
{Means}%
}{%
{Means}%
}{%
{\protect \APACyear {1972}}%
}]{%
Means72}
\APACinsertmetastar {%
Means72}%
\begin{APACrefauthors}%
{Means}, J\BPBI D.%
\end{APACrefauthors}%
\unskip\
\newblock
\APACrefYearMonthDay{1972}{}{}.
\newblock
{\BBOQ}\APACrefatitle {{Use of the three-dimensional covariance matrix in analyzing the polarization properties of plane waves}} {{Use of the three-dimensional covariance matrix in analyzing the polarization properties of plane waves}}.{\BBCQ}
\newblock
\APACjournalVolNumPages{\jgr}{77}{}{5551}.
\newblock
\begin{APACrefDOI} \doi{10.1029/JA077i028p05551} \end{APACrefDOI}
\PrintBackRefs{\CurrentBib}

\bibitem [\protect \citeauthoryear {%
{Min}%
\ \protect \BOthers {.}}{%
{Min}%
\ \protect \BOthers {.}}{%
{\protect \APACyear {2015}}%
}]{%
Min15:emic}
\APACinsertmetastar {%
Min15:emic}%
\begin{APACrefauthors}%
{Min}, K.%
, {Liu}, K.%
, {Bonnell}, J\BPBI W.%
, {Breneman}, A\BPBI W.%
, {Denton}, R\BPBI E.%
, {Funsten}, H\BPBI O.%
\BDBL {}{Wygant}, J\BPBI R.%
\end{APACrefauthors}%
\unskip\
\newblock
\APACrefYearMonthDay{2015}{{\APACmonth{04}}}{}.
\newblock
{\BBOQ}\APACrefatitle {{Study of EMIC wave excitation using direct ion measurements}} {{Study of EMIC wave excitation using direct ion measurements}}.{\BBCQ}
\newblock
\APACjournalVolNumPages{Journal of Geophysical Research (Space Physics)}{120}{4}{2702-2719}.
\newblock
\begin{APACrefDOI} \doi{10.1002/2014JA020717} \end{APACrefDOI}
\PrintBackRefs{\CurrentBib}

\bibitem [\protect \citeauthoryear {%
{Mourenas}%
, {Artemyev}%
, {Agapitov}%
, {Krasnoselskikh}%
\BCBL {}\ \BBA {} {Mozer}%
}{%
{Mourenas}%
\ \protect \BOthers {.}}{%
{\protect \APACyear {2015}}%
}]{%
Mourenas15}
\APACinsertmetastar {%
Mourenas15}%
\begin{APACrefauthors}%
{Mourenas}, D.%
, {Artemyev}, A\BPBI V.%
, {Agapitov}, O\BPBI V.%
, {Krasnoselskikh}, V.%
\BCBL {}\ \BBA {} {Mozer}, F\BPBI S.%
\end{APACrefauthors}%
\unskip\
\newblock
\APACrefYearMonthDay{2015}{}{}.
\newblock
{\BBOQ}\APACrefatitle {{Very oblique whistler generation by low-energy electron streams}} {{Very oblique whistler generation by low-energy electron streams}}.{\BBCQ}
\newblock
\APACjournalVolNumPages{\jgr}{120}{}{3665--3683}.
\newblock
\begin{APACrefDOI} \doi{10.1002/2015JA021135} \end{APACrefDOI}
\PrintBackRefs{\CurrentBib}

\bibitem [\protect \citeauthoryear {%
{Ni}%
\ \protect \BOthers {.}}{%
{Ni}%
\ \protect \BOthers {.}}{%
{\protect \APACyear {2016}}%
}]{%
Ni16:ssr}
\APACinsertmetastar {%
Ni16:ssr}%
\begin{APACrefauthors}%
{Ni}, B.%
, {Thorne}, R\BPBI M.%
, {Zhang}, X.%
, {Bortnik}, J.%
, {Pu}, Z.%
, {Xie}, L.%
\BDBL {}{Gu}, X.%
\end{APACrefauthors}%
\unskip\
\newblock
\APACrefYearMonthDay{2016}{{\APACmonth{04}}}{}.
\newblock
{\BBOQ}\APACrefatitle {{Origins of the Earth's Diffuse Auroral Precipitation}} {{Origins of the Earth's Diffuse Auroral Precipitation}}.{\BBCQ}
\newblock
\APACjournalVolNumPages{\ssr}{200}{}{205-259}.
\newblock
\begin{APACrefDOI} \doi{10.1007/s11214-016-0234-7} \end{APACrefDOI}
\PrintBackRefs{\CurrentBib}

\bibitem [\protect \citeauthoryear {%
{Nishimura}%
\ \protect \BOthers {.}}{%
{Nishimura}%
\ \protect \BOthers {.}}{%
{\protect \APACyear {2013}}%
}]{%
Nishimura13:density}
\APACinsertmetastar {%
Nishimura13:density}%
\begin{APACrefauthors}%
{Nishimura}, Y.%
, {Bortnik}, J.%
, {Li}, W.%
, {Thorne}, R\BPBI M.%
, {Ni}, B.%
, {Lyons}, L\BPBI R.%
\BDBL {}{Auster}, U.%
\end{APACrefauthors}%
\unskip\
\newblock
\APACrefYearMonthDay{2013}{Feb}{}.
\newblock
{\BBOQ}\APACrefatitle {{Structures of dayside whistler-mode waves deduced from conjugate diffuse aurora}} {{Structures of dayside whistler-mode waves deduced from conjugate diffuse aurora}}.{\BBCQ}
\newblock
\APACjournalVolNumPages{Journal of Geophysical Research (Space Physics)}{118}{2}{664-673}.
\newblock
\begin{APACrefDOI} \doi{10.1029/2012JA018242} \end{APACrefDOI}
\PrintBackRefs{\CurrentBib}

\bibitem [\protect \citeauthoryear {%
{Omura}%
, {Ashour-Abdalla}%
, {Gendrin}%
\BCBL {}\ \BBA {} {Quest}%
}{%
{Omura}%
\ \protect \BOthers {.}}{%
{\protect \APACyear {1985}}%
}]{%
Omura85}
\APACinsertmetastar {%
Omura85}%
\begin{APACrefauthors}%
{Omura}, Y.%
, {Ashour-Abdalla}, M.%
, {Gendrin}, R.%
\BCBL {}\ \BBA {} {Quest}, K.%
\end{APACrefauthors}%
\unskip\
\newblock
\APACrefYearMonthDay{1985}{{\APACmonth{09}}}{}.
\newblock
{\BBOQ}\APACrefatitle {{Heating of thermal helium in the equatorial magnetosphere: A simulation study}} {{Heating of thermal helium in the equatorial magnetosphere: A simulation study}}.{\BBCQ}
\newblock
\APACjournalVolNumPages{\jgr}{90}{A9}{8281-8292}.
\newblock
\begin{APACrefDOI} \doi{10.1029/JA090iA09p08281} \end{APACrefDOI}
\PrintBackRefs{\CurrentBib}

\bibitem [\protect \citeauthoryear {%
{Rauch}%
\ \BBA {} {Roux}%
}{%
{Rauch}%
\ \BBA {} {Roux}%
}{%
{\protect \APACyear {1982}}%
}]{%
Rauch&Roux82}
\APACinsertmetastar {%
Rauch&Roux82}%
\begin{APACrefauthors}%
{Rauch}, J\BPBI L.%
\BCBT {}\ \BBA {} {Roux}, A.%
\end{APACrefauthors}%
\unskip\
\newblock
\APACrefYearMonthDay{1982}{{\APACmonth{10}}}{}.
\newblock
{\BBOQ}\APACrefatitle {{Ray tracing of ULF waves in a multicomponent magnetospheric plasma: Consequences for the generation mechanism of ion cyclotron waves}} {{Ray tracing of ULF waves in a multicomponent magnetospheric plasma: Consequences for the generation mechanism of ion cyclotron waves}}.{\BBCQ}
\newblock
\APACjournalVolNumPages{\jgr}{87}{A10}{8191-8198}.
\newblock
\begin{APACrefDOI} \doi{10.1029/JA087iA10p08191} \end{APACrefDOI}
\PrintBackRefs{\CurrentBib}

\bibitem [\protect \citeauthoryear {%
{Silin}%
, {Mann}%
, {Sydora}%
, {Summers}%
\BCBL {}\ \BBA {} {Mace}%
}{%
{Silin}%
\ \protect \BOthers {.}}{%
{\protect \APACyear {2011}}%
}]{%
Silin11}
\APACinsertmetastar {%
Silin11}%
\begin{APACrefauthors}%
{Silin}, I.%
, {Mann}, I\BPBI R.%
, {Sydora}, R\BPBI D.%
, {Summers}, D.%
\BCBL {}\ \BBA {} {Mace}, R\BPBI L.%
\end{APACrefauthors}%
\unskip\
\newblock
\APACrefYearMonthDay{2011}{May}{}.
\newblock
{\BBOQ}\APACrefatitle {{Warm plasma effects on electromagnetic ion cyclotron wave MeV electron interactions in the magnetosphere}} {{Warm plasma effects on electromagnetic ion cyclotron wave MeV electron interactions in the magnetosphere}}.{\BBCQ}
\newblock
\APACjournalVolNumPages{Journal of Geophysical Research (Space Physics)}{116}{A5}{A05215}.
\newblock
\begin{APACrefDOI} \doi{10.1029/2010JA016398} \end{APACrefDOI}
\PrintBackRefs{\CurrentBib}

\bibitem [\protect \citeauthoryear {%
{Smirnov}%
\ \BBA {} {Frank-Kamenestki{\v{i}}}%
}{%
{Smirnov}%
\ \BBA {} {Frank-Kamenestki{\v{i}}}%
}{%
{\protect \APACyear {1968}}%
}]{%
Smirnov&FrankKamenestkii68}
\APACinsertmetastar {%
Smirnov&FrankKamenestkii68}%
\begin{APACrefauthors}%
{Smirnov}, Y\BPBI N.%
\BCBT {}\ \BBA {} {Frank-Kamenestki{\v{i}}}, D\BPBI A.%
\end{APACrefauthors}%
\unskip\
\newblock
\APACrefYearMonthDay{1968}{{\APACmonth{03}}}{}.
\newblock
{\BBOQ}\APACrefatitle {{Nonlinearity and Parametric Resonance in a Plasma}} {{Nonlinearity and Parametric Resonance in a Plasma}}.{\BBCQ}
\newblock
\APACjournalVolNumPages{Soviet Journal of Experimental and Theoretical Physics}{26}{}{627}.
\PrintBackRefs{\CurrentBib}

\bibitem [\protect \citeauthoryear {%
{Stix}%
}{%
{Stix}%
}{%
{\protect \APACyear {1962}}%
}]{%
bookStix62}
\APACinsertmetastar {%
bookStix62}%
\begin{APACrefauthors}%
{Stix}, T\BPBI H.%
\end{APACrefauthors}%
\unskip\
\newblock
\APACrefYear{1962}.
\newblock
\APACrefbtitle {{The Theory of Plasma Waves}} {{The Theory of Plasma Waves}}.
\PrintBackRefs{\CurrentBib}

\bibitem [\protect \citeauthoryear {%
{Terasawa}%
\ \BBA {} {Matsukiyo}%
}{%
{Terasawa}%
\ \BBA {} {Matsukiyo}%
}{%
{\protect \APACyear {2012}}%
}]{%
Terasawa&Matsukiyo12}
\APACinsertmetastar {%
Terasawa&Matsukiyo12}%
\begin{APACrefauthors}%
{Terasawa}, T.%
\BCBT {}\ \BBA {} {Matsukiyo}, S.%
\end{APACrefauthors}%
\unskip\
\newblock
\APACrefYearMonthDay{2012}{{\APACmonth{11}}}{}.
\newblock
{\BBOQ}\APACrefatitle {{Cyclotron Resonant Interactions in Cosmic Particle Accelerators}} {{Cyclotron Resonant Interactions in Cosmic Particle Accelerators}}.{\BBCQ}
\newblock
\APACjournalVolNumPages{\ssr}{173}{}{623-640}.
\newblock
\begin{APACrefDOI} \doi{10.1007/s11214-012-9878-0} \end{APACrefDOI}
\PrintBackRefs{\CurrentBib}

\bibitem [\protect \citeauthoryear {%
{Thorne}%
\ \BBA {} {Horne}%
}{%
{Thorne}%
\ \BBA {} {Horne}%
}{%
{\protect \APACyear {1992}}%
}]{%
Thorne&Horne92}
\APACinsertmetastar {%
Thorne&Horne92}%
\begin{APACrefauthors}%
{Thorne}, R\BPBI M.%
\BCBT {}\ \BBA {} {Horne}, R\BPBI B.%
\end{APACrefauthors}%
\unskip\
\newblock
\APACrefYearMonthDay{1992}{{\APACmonth{02}}}{}.
\newblock
{\BBOQ}\APACrefatitle {{The contribution of ion-cyclotron waves to electron heating and SAR-arc excitation near the storm-time plasmapause}} {{The contribution of ion-cyclotron waves to electron heating and SAR-arc excitation near the storm-time plasmapause}}.{\BBCQ}
\newblock
\APACjournalVolNumPages{\grl}{19}{4}{417-420}.
\newblock
\begin{APACrefDOI} \doi{10.1029/92GL00089} \end{APACrefDOI}
\PrintBackRefs{\CurrentBib}

\bibitem [\protect \citeauthoryear {%
{Thorne}%
\ \BBA {} {Horne}%
}{%
{Thorne}%
\ \BBA {} {Horne}%
}{%
{\protect \APACyear {1997}}%
}]{%
Thorne&Horne97}
\APACinsertmetastar {%
Thorne&Horne97}%
\begin{APACrefauthors}%
{Thorne}, R\BPBI M.%
\BCBT {}\ \BBA {} {Horne}, R\BPBI B.%
\end{APACrefauthors}%
\unskip\
\newblock
\APACrefYearMonthDay{1997}{{\APACmonth{07}}}{}.
\newblock
{\BBOQ}\APACrefatitle {{Modulation of electromagnetic ion cyclotron instability due to interaction with ring current O$^{+}$ during magnetic storms}} {{Modulation of electromagnetic ion cyclotron instability due to interaction with ring current O$^{+}$ during magnetic storms}}.{\BBCQ}
\newblock
\APACjournalVolNumPages{\jgr}{102}{A7}{14155-14164}.
\newblock
\begin{APACrefDOI} \doi{10.1029/96JA04019} \end{APACrefDOI}
\PrintBackRefs{\CurrentBib}

\bibitem [\protect \citeauthoryear {%
{Thorne}%
, {Ni}%
, {Tao}%
, {Horne}%
\BCBL {}\ \BBA {} {Meredith}%
}{%
{Thorne}%
\ \protect \BOthers {.}}{%
{\protect \APACyear {2010}}%
}]{%
Thorne10:Nature}
\APACinsertmetastar {%
Thorne10:Nature}%
\begin{APACrefauthors}%
{Thorne}, R\BPBI M.%
, {Ni}, B.%
, {Tao}, X.%
, {Horne}, R\BPBI B.%
\BCBL {}\ \BBA {} {Meredith}, N\BPBI P.%
\end{APACrefauthors}%
\unskip\
\newblock
\APACrefYearMonthDay{2010}{{\APACmonth{10}}}{}.
\newblock
{\BBOQ}\APACrefatitle {{Scattering by chorus waves as the dominant cause of diffuse auroral precipitation}} {{Scattering by chorus waves as the dominant cause of diffuse auroral precipitation}}.{\BBCQ}
\newblock
\APACjournalVolNumPages{\nat}{467}{}{943-946}.
\newblock
\begin{APACrefDOI} \doi{10.1038/nature09467} \end{APACrefDOI}
\PrintBackRefs{\CurrentBib}

\bibitem [\protect \citeauthoryear {%
{Tsai}%
, {Artemyev}%
, {Angelopoulos}%
\BCBL {}\ \BBA {} {Zhang}%
}{%
{Tsai}%
\ \protect \BOthers {.}}{%
{\protect \APACyear {2023}}%
}]{%
Tsai23}
\APACinsertmetastar {%
Tsai23}%
\begin{APACrefauthors}%
{Tsai}, E.%
, {Artemyev}, A.%
, {Angelopoulos}, V.%
\BCBL {}\ \BBA {} {Zhang}, X\BHBI J.%
\end{APACrefauthors}%
\unskip\
\newblock
\APACrefYearMonthDay{2023}{{\APACmonth{08}}}{}.
\newblock
{\BBOQ}\APACrefatitle {{Investigating Whistler-Mode Wave Intensity Along Field Lines Using Electron Precipitation Measurements}} {{Investigating Whistler-Mode Wave Intensity Along Field Lines Using Electron Precipitation Measurements}}.{\BBCQ}
\newblock
\APACjournalVolNumPages{Journal of Geophysical Research (Space Physics)}{128}{8}{e2023JA031578}.
\newblock
\begin{APACrefDOI} \doi{10.1029/2023JA031578} \end{APACrefDOI}
\PrintBackRefs{\CurrentBib}

\bibitem [\protect \citeauthoryear {%
{Ukhorskiy}%
\ \protect \BOthers {.}}{%
{Ukhorskiy}%
\ \protect \BOthers {.}}{%
{\protect \APACyear {2018}}%
}]{%
Ukhorskiy18:DF}
\APACinsertmetastar {%
Ukhorskiy18:DF}%
\begin{APACrefauthors}%
{Ukhorskiy}, A\BPBI Y.%
, {Sorathia}, K\BPBI A.%
, {Merkin}, V\BPBI G.%
, {Sitnov}, M\BPBI I.%
, {Mitchell}, D\BPBI G.%
\BCBL {}\ \BBA {} {Gkioulidou}, M.%
\end{APACrefauthors}%
\unskip\
\newblock
\APACrefYearMonthDay{2018}{Jul}{}.
\newblock
{\BBOQ}\APACrefatitle {{Ion Trapping and Acceleration at Dipolarization Fronts: High-Resolution MHD and Test-Particle Simulations}} {{Ion Trapping and Acceleration at Dipolarization Fronts: High-Resolution MHD and Test-Particle Simulations}}.{\BBCQ}
\newblock
\APACjournalVolNumPages{Journal of Geophysical Research (Space Physics)}{123}{7}{5580-5589}.
\newblock
\begin{APACrefDOI} \doi{10.1029/2018JA025370} \end{APACrefDOI}
\PrintBackRefs{\CurrentBib}

\bibitem [\protect \citeauthoryear {%
{Usanova}%
}{%
{Usanova}%
}{%
{\protect \APACyear {2021}}%
}]{%
Usanova21}
\APACinsertmetastar {%
Usanova21}%
\begin{APACrefauthors}%
{Usanova}, M\BPBI E.%
\end{APACrefauthors}%
\unskip\
\newblock
\APACrefYearMonthDay{2021}{{\APACmonth{09}}}{}.
\newblock
{\BBOQ}\APACrefatitle {{Energy Exchange Between Electromagnetic Ion Cyclotron (EMIC) Waves and Thermal Plasma: From Theory to Observations}} {{Energy Exchange Between Electromagnetic Ion Cyclotron (EMIC) Waves and Thermal Plasma: From Theory to Observations}}.{\BBCQ}
\newblock
\APACjournalVolNumPages{Frontiers in Astronomy and Space Sciences}{8}{}{150}.
\newblock
\begin{APACrefDOI} \doi{10.3389/fspas.2021.744344} \end{APACrefDOI}
\PrintBackRefs{\CurrentBib}

\bibitem [\protect \citeauthoryear {%
{Usanova}%
, {Mann}%
\BCBL {}\ \BBA {} {Darrouzet}%
}{%
{Usanova}%
\ \protect \BOthers {.}}{%
{\protect \APACyear {2016}}%
}]{%
Usanova&Mann16}
\APACinsertmetastar {%
Usanova&Mann16}%
\begin{APACrefauthors}%
{Usanova}, M\BPBI E.%
, {Mann}, I\BPBI R.%
\BCBL {}\ \BBA {} {Darrouzet}, F.%
\end{APACrefauthors}%
\unskip\
\newblock
\APACrefYearMonthDay{2016}{{\APACmonth{02}}}{}.
\newblock
{\BBOQ}\APACrefatitle {{EMIC Waves in the Inner Magnetosphere}} {{EMIC Waves in the Inner Magnetosphere}}.{\BBCQ}
\newblock
\APACjournalVolNumPages{Washington DC American Geophysical Union Geophysical Monograph Series}{216}{}{65-78}.
\newblock
\begin{APACrefDOI} \doi{10.1002/9781119055006.ch5} \end{APACrefDOI}
\PrintBackRefs{\CurrentBib}

\bibitem [\protect \citeauthoryear {%
B.~{Wang}%
, {Li}%
, {Huang}%
\BCBL {}\ \BBA {} {Zhang}%
}{%
B.~{Wang}%
\ \protect \BOthers {.}}{%
{\protect \APACyear {2019}}%
}]{%
Wang19:emic}
\APACinsertmetastar {%
Wang19:emic}%
\begin{APACrefauthors}%
{Wang}, B.%
, {Li}, P.%
, {Huang}, J.%
\BCBL {}\ \BBA {} {Zhang}, B.%
\end{APACrefauthors}%
\unskip\
\newblock
\APACrefYearMonthDay{2019}{{\APACmonth{04}}}{}.
\newblock
{\BBOQ}\APACrefatitle {{Nonlinear Landau resonance between EMIC waves and cold electrons in the inner magnetosphere}} {{Nonlinear Landau resonance between EMIC waves and cold electrons in the inner magnetosphere}}.{\BBCQ}
\newblock
\APACjournalVolNumPages{Physics of Plasmas}{26}{4}{042903}.
\newblock
\begin{APACrefDOI} \doi{10.1063/1.5088374} \end{APACrefDOI}
\PrintBackRefs{\CurrentBib}

\bibitem [\protect \citeauthoryear {%
Q.~{Wang}%
\ \protect \BOthers {.}}{%
Q.~{Wang}%
\ \protect \BOthers {.}}{%
{\protect \APACyear {2018}}%
}]{%
Wang18:bounce}
\APACinsertmetastar {%
Wang18:bounce}%
\begin{APACrefauthors}%
{Wang}, Q.%
, {Fu}, S.%
, {Ni}, B.%
, {Cao}, X.%
, {Gu}, X.%
\BCBL {}\ \BBA {} {Huang}, H.%
\end{APACrefauthors}%
\unskip\
\newblock
\APACrefYearMonthDay{2018}{Aug}{}.
\newblock
{\BBOQ}\APACrefatitle {{Bounce resonance scattering of ring current electrons by H$^{+}$ band EMIC waves}} {{Bounce resonance scattering of ring current electrons by H$^{+}$ band EMIC waves}}.{\BBCQ}
\newblock
\APACjournalVolNumPages{Physics of Plasmas}{25}{8}{082903}.
\newblock
\begin{APACrefDOI} \doi{10.1063/1.5043522} \end{APACrefDOI}
\PrintBackRefs{\CurrentBib}

\bibitem [\protect \citeauthoryear {%
Yue%
\ \protect \BOthers {.}}{%
Yue%
\ \protect \BOthers {.}}{%
{\protect \APACyear {2017}}%
}]{%
Yue17:ions}
\APACinsertmetastar {%
Yue17:ions}%
\begin{APACrefauthors}%
Yue, C.%
, Bortnik, J.%
, Thorne, R\BPBI M.%
, Ma, Q.%
, An, X.%
, Chappell, C\BPBI R.%
\BDBL {}Kletzing, C\BPBI A.%
\end{APACrefauthors}%
\unskip\
\newblock
\APACrefYearMonthDay{2017}{}{}.
\newblock
{\BBOQ}\APACrefatitle {The characteristic pitch angle distributions of 1 eV to 600 keV protons near the equator based on Van Allen Probes observations} {The characteristic pitch angle distributions of 1 ev to 600 kev protons near the equator based on van allen probes observations}.{\BBCQ}
\newblock
\APACjournalVolNumPages{\jgr}{}{}{}.
\newblock
\begin{APACrefURL} \url{http://dx.doi.org/10.1002/2017JA024421} \end{APACrefURL}
\newblock
\begin{APACrefDOI} \doi{10.1002/2017JA024421} \end{APACrefDOI}
\PrintBackRefs{\CurrentBib}

\bibitem [\protect \citeauthoryear {%
{Yue}%
\ \protect \BOthers {.}}{%
{Yue}%
\ \protect \BOthers {.}}{%
{\protect \APACyear {2019}}%
}]{%
Yue19:emic}
\APACinsertmetastar {%
Yue19:emic}%
\begin{APACrefauthors}%
{Yue}, C.%
, {Jun}, C\BHBI W.%
, {Bortnik}, J.%
, {An}, X.%
, {Ma}, Q.%
, {Reeves}, G\BPBI D.%
\BDBL {}{Kletzing}, C\BPBI A.%
\end{APACrefauthors}%
\unskip\
\newblock
\APACrefYearMonthDay{2019}{Apr}{}.
\newblock
{\BBOQ}\APACrefatitle {{The Relationship Between EMIC Wave Properties and Proton Distributions Based on Van Allen Probes Observations}} {{The Relationship Between EMIC Wave Properties and Proton Distributions Based on Van Allen Probes Observations}}.{\BBCQ}
\newblock
\APACjournalVolNumPages{\grl}{46}{8}{4070-4078}.
\newblock
\begin{APACrefDOI} \doi{10.1029/2019GL082633} \end{APACrefDOI}
\PrintBackRefs{\CurrentBib}

\end{thebibliography}

\end{document}